\documentclass[useAMS,usenatbib]{mn2e}

\usepackage[british]{babel}
\usepackage{graphicx,amssymb,amsmath,epsfig,rotating,fleqn}
\usepackage{color}
\usepackage{float}
\usepackage{rotating}
\usepackage{tikz}
\usepackage{lscape}

\DeclareRobustCommand\reddot{\tikz \filldraw[fill=red,draw=black] circle (2pt);}
\DeclareRobustCommand\bluedot{\tikz \filldraw[fill=blue,draw=black] circle (2pt);}
\DeclareRobustCommand\whitedot{\tikz \filldraw[fill=white,draw=black] circle (2pt);}

\title
  [Self-clustering of cosmic rays]
  {A Bayesian self-clustering analysis of the highest energy cosmic rays detected by the Pierre Auger Observatory}
\author
  [A.\ Khanin \& D.\ J.\ Mortlock]
  {Alexander Khanin$^{1}$\thanks{E-mail: ak2008@imperial.ac.uk}
  and 
  Daniel J.\ Mortlock$^{1,2}$
\vspace{3mm}\\
$^1$Astrophysics Group, Imperial College London, Blackett Laboratory,
  Prince Consort Road, London SW7 2AZ, U.K. \\
$^2$Department of Mathematics, Imperial College London, London SW7 2AZ, U.K.
   }
\begin{document}

\date{Accepted 2014 ?????? ??. 
  Received 2014 ?????? ??; in original form 2014 ???????? ??}

\pagerange{\pageref{firstpage}--\pageref{lastpage}} \pubyear{2014}

\maketitle

\label{firstpage}

\begin{abstract}
Cosmic rays (CRs) are protons and atomic nuclei that flow into our Solar system and reach the Earth with energies of up to $\sim10^{21}\,\rm{eV}$. The sources of ultra-high energy cosmic rays (UHECRs) with $E \gtrsim 10^{19}\,\rm{eV}$ remain unknown, although there are theoretical reasons to think that at least some come from active galactic nuclei (AGNs). One way to assess the different hypotheses is by analysing the arrival directions of UHECRs, in particular their self-clustering. We have developed a fully Bayesian approach to analyzing the self-clustering of points on the sphere, which we apply to the UHECR arrival directions. The analysis is based on a multi-step approach that enables the application of Bayesian model comparison to cases with weak prior information. We have applied this approach to the 69 highest energy events recorded by the Pierre Auger Observatory (PAO), which is the largest current UHECR data set. We do not detect self-clustering, but simulations show that this is consistent with the AGN-sourced model for a data set of this size. Data sets of several hundred UHECRs would be sufficient to detect clustering in the AGN model. Samples of this magnitude are expected to be produced by future experiments, such as the Japanese Experiment Module Extreme Universe Space Observatory (JEM-EUSO).
\end{abstract}



\section{Introduction}
\label{section:intro}

Cosmic rays (CRs) are high-energy particles that flow into our Solar system and reach the Earth. They consist mainly of protons and atomic nuclei, and have energies in the range $10^{9}\,{\rm eV}$ to $10^{21}\,{\rm eV}$, which makes them the most energetic particles observed in nature (see e.g.\ \citealt{Stanev2011} for a review). A number of open issues remain in this field, especially with respect to ultra-high energy cosmic rays (UHECRs) with arrival energies $E_{\rm{arr}} \gtrsim 10^{19}\,{\rm eV}$. In particular, no consensus has been reached on the sources of UHECRs. A number of candidates, such as active galactic nuclei (AGNs) and pulsars have been proposed, but lack empirical verification.

The strongest demonstration of the origin of the UHECRs would be if they could be associated with their progenitors, something which is made plausible by the fact that the most energetic CRs can only travel for cosmologically short distances before losing energy. UHECRs with energies of $E \ga 5\times 10^{19}\, {\rm eV}$ scatter off the cosmic microwave background (CMB) radiation via the Greisen-Zatsepin-Kuzmin (GZK) effect (\citealt{Greisen66}, \citealt{ZK66}). The resultant energy loss is very significant: the mean free path of the  GZK effect at high energies is a few ${\rm Mpc}$ and the energy loss in each collision is 20-50 \% depending on energy (\citealt{Stanev2009}). The GZK effect is expected to cause an abrupt cutoff in the flux of UHECRs at $\sim$ $4 \times 10^{19}\, {\rm eV}$, for which there has been observational support (\citealt{CutoffPAO,CutoffHiRes}). UHECRs that arrive at Earth with energies above the GZK limit can only have come from within a limited radius (the GZK horizon) of $\sim 100 \,{\rm Mpc}$. 

Due to the low flux, the number of detected UHECRs is small: the largest currently available sample is the 69 events with $E \geq 5.5\times 10^{19}\, {\rm eV}$ recorded by the Pierre Auger Observatory (PAO) between 2004 January 1 and 2009 December 31 (\citealt{PAO2010}). The low number of events is the main reason why any hypothesis about the sources is difficult to investigate. 

Another difficulty is that CRs are charged particles, and so are deflected by magnetic fields. The deflection due to the extra-Galactic magnetic fields is expected to be $\sim 2$ to  $\sim 10$ deg for the highest energy CRs (e.g. \citealt{Medina1998,Sigl2004,Dolag2005}). This complicates the study of UHECR origins because it becomes becomes difficult to directly link arrival directions with possible sources.

Nevertheless, a number of attempts have been made to find a correlation between the arrival directions of UHECRs and catalogues of potential sources, although no clear consensus has yet been reached.  The Pierre Auger Collaboration reported a strong correlation between the arrival directions of UHECRs with energies $E \ge 5.7 \times 10^{19}\,{\rm eV}$ and the positions of nearby AGNs (\citealt{PAO2007}). The result is supported by Yakutsk data (\citealt{Ivanov2009}), but not by HiRes (\citealt{Abbasi2008}) or Telescope Array (\citealt{AbuZayyad2012}). A more recent analysis of a larger PAO UHECR sample has shown a much weaker correlation than before (\citealt{PAO2010}). 

All attempts to associate UHECRs with specific sources are hampered to some degree by large magnetic deflections, possibly transient sources and incomplete catalogues. An alternative approach is based on the idea that if the UHECR sources are distributed inhomogeneously inside the GZK horizon, it should be possible to detect a self-clustering in the UHECR arrival directions, independent of any source catalogue. Examples of such work include \cite{Domenico2011} and \cite{PAO2012}. In \cite{PAO2012}, the Pierre Auger Collaboration studied the self-clustering using three statistical methods based on correlation functions (two methods based on the 2-point correlation function, one method based on a 3-point correlation function, developed by \citealt{Ave2009}). No strong evidence of non-uniformity was found based on the $p$-values obtained under the null hypothesis of no clustering. The interpretation of $p$-values is, however, known to be problematic as they have no quantitative link to the (posterior) probability that the null hypothesis is correct (see e.g.\ \citealt{Berger_Delampady1987}).  

Whereas $p$-values are probabilities conditional on the null hypothesis, what is needed is a method of calculating the probability that the null hypothesis is correct. \cite{Cox1946} proved that Bayesian inference is the only self-consistent method to make probabilistic statements about models based on observations, and Bayesian methods have previously been used to assess whether UHECRs originate from AGNs \citep{Watson_etal2011,Soiaporn2012}.  

In this paper we present a Bayesian analysis of the self-clustering of the PAO UHECRs. The Bayesian method for assessing non-uniformity is explained in Section~\ref{sec:StatMeth}. In Section~\ref{sec:ApplicationSimul}, the effectiveness of the method is discussed, based on tests of the method on simulated mock UHECR catalogues. The application of the method to data from PAO is discussed in Section~\ref{sec:ApplicationPAO}. Our conclusions are summarized in Section~\ref{sec:Conc}.
\vspace{1mm}
\section{Statistical formalism}
\label{sec:StatMeth}

Our primary aim here is to assess whether there is evidence that the distribution of UHECR arrival directions is anisotropic. We do this by using Bayesian inference in the context of two models: a uniform model, $M_{\rm u}$, which would be the null hypothesis in a classical hypothesis test; and a non-uniform model, $M_{\rm n}$, as yet unspecified.  The posterior probability of the non-uniform model, conditional on data in the form of $N$ UHECR arrival directions $\{ \bmath{r}_i \}$ (where $i \in \{1, 2, \ldots, N\}$), is given by Bayes's theorem as 
\[
{\rm Pr}(M_{\rm n} | \{ \bmath{r}_i \} ) 
\]
\begin{equation}
\label{eq:BayesTheorem}
\mbox{} = \frac{{\rm Pr}(M_{\rm n}) \, {\rm Pr}(\{ \bmath{r}_i \}|M_{\rm n})}{ {\rm Pr}(M_{\rm u})\,{\rm Pr}(\{ \bmath{r}_i \}|M_{\rm u}) + {\rm Pr}(M_{\rm n})\,{\rm Pr}(\{ \bmath{r}_i \}|M_{\rm n}) },
\end{equation}
where ${\rm Pr}(M_{\rm u})$ and ${\rm Pr}(M_{\rm n})$ are the prior probabilities of the two models, and  ${\rm Pr}(\{ \bmath{r}_i \}|M_{\rm u})$ and  ${\rm Pr}(\{ \bmath{r}_i \}|M_{\rm n})$ are the probabilities of the data under each of the models (i.e.\ the likelihoods).
With just two models, it is convenient to work with the ratio of the posterior probabilities, given by
\begin{equation}
    \label{eq:BayesTheorem2}
\frac{{\rm Pr}(M_{\rm n} | \{ \bmath{r}_i \} )}{{\rm Pr}(M_{\rm u} | \{ \bmath{r}_i \} )} = \frac{{\rm Pr}(M_{\rm n} )}{{\rm Pr}(M_{\rm u})}B  ,
\end{equation}
where 
\begin{equation}
 \label{eq:BayesFactor}
  B = \frac{{\rm Pr}(\{ \bmath{r}_i \}|M_{\rm n})}{{\rm Pr}(\{ \bmath{r}_i \}|M_{\rm u})}
\end{equation}
is the Bayes factor. In the convention adopted here, models $M_{\rm u}$ and $M_{\rm n}$ are favoured by small and large values of $B$, respectively.

If a model $M$ has an unspecified parameter $\theta$, then ${\rm Pr}(\{ \bmath{r}_i \}|M)$ is the marginal likelihood\footnote{The marginal likelihood is sometimes referred to as the model-averaged likelihood or, particularly in astronomy, as the (Bayesian) evidence.}, which is given by  
\begin{equation}
    \label{eq:Evidence}
{\rm Pr}(\{ \bmath{r}_i \}|M) = \int_{-\infty}^\infty {\rm Pr}(\theta |M) \, {\rm Pr}(\{ \bmath{r}_i \}|\theta ,M) \,{\rm d}\theta  ,
\end{equation}
where ${\rm Pr}(\{ \bmath{r}_i \}|\theta ,M)$ is the probability of the data for a given value of $\theta$ and ${\rm Pr}(\theta |M)$ is the prior distribution of the parameter. This distribution must be fully specified and unit-normalized, otherwise the resultant value of ${\rm Pr}(\{ \bmath{r}_i \}|M)$ is meaningless (\citealt{Jeffreys1961}).

The next task is to specify the two models to be compared and to evaluate the marginal likelihoods for both.  The null hypothesis represented by the uniform model (Section~\ref{sec:uniform}) is unambiguous and yields the marginal likelihood given in Equation~\ref{eq:BayesFactorU}; the alternative non-uniform model (Section~\ref{sec:nonuniform}) is more complicated and is derived from a subset of the data, eventually yielding the marginal likelihood given in Equation~\ref{eq:ProbGen}.  This requirement means that both marginal likelihoods are evaluated only for the remaining data that was not used to obtain the non-uniform model.

\subsection{Uniform model}
\label{sec:uniform}
In the uniform model, $M_{\rm u}$, the probability that a UHECR arrives from direction $\bmath{r}$ is constant at ${\rm Pr}(\bmath{r}|M_{\rm u})=1/(4\pi)$. Hence, the marginal likelihood for a test sample of $N_{\rm t}$ UHECRs with arrival direction $\{\bmath{r}_t\}$ (with $t \in \{1, 2, \ldots, N_{\rm t}\}$) is given by 
\begin{equation}
 \label{eq:BayesFactorU}
 {\rm Pr}(\{\bmath{r}_t\}|M_{\rm u})=  \frac{1}{(4\rmn{\pi})^{N_{\rm t}}} .
\end{equation}
This simple expression is, however, valid only in the case of uniform exposure; if the exposure is non-uniform, as is always the case for real experiments, it must be modified as described in Section~\ref{sec:exposure}.

\clearpage

\begin{landscape}
\begin{figure}
\vspace{1.5cm}
\begin{center}$
\arraycolsep=0.01pt\def\arraystretch{0.01}
\begin{array}{cccc}
(1) & Uniform  & Three \, sources & AGN \, sources\\
(2) & \includegraphics[width=75mm]{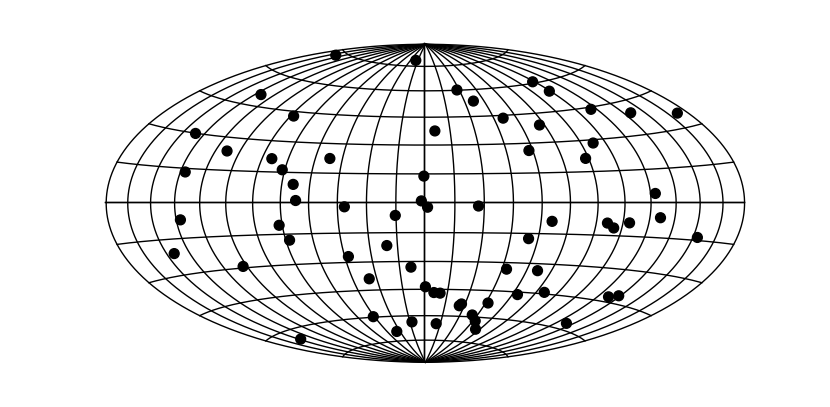}&\includegraphics[width=75mm]{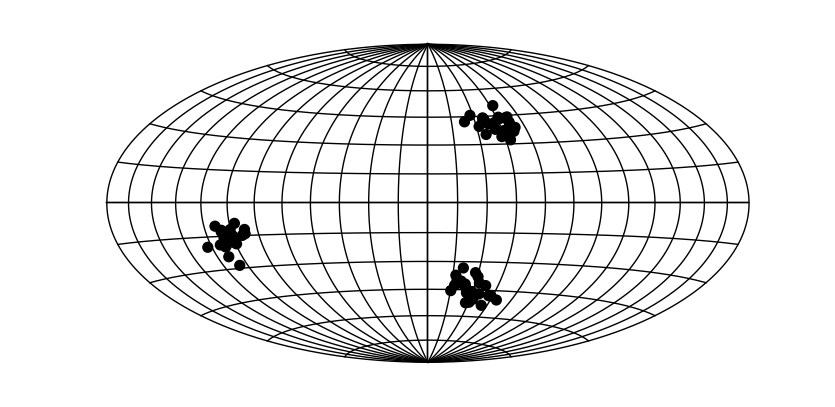}&\includegraphics[width=75mm]{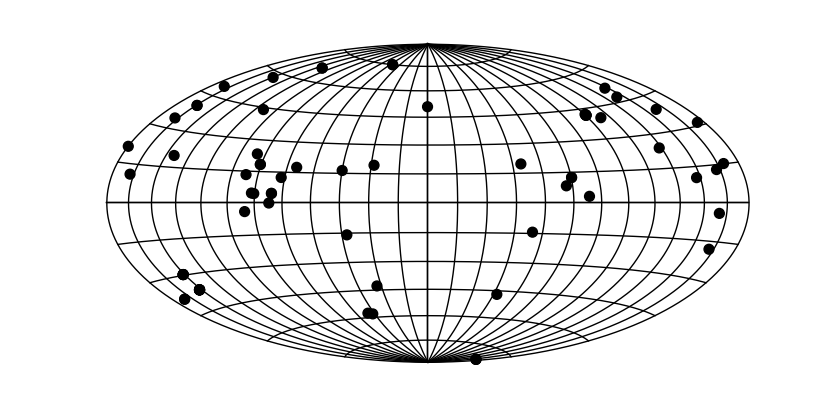}\\

(3)& \includegraphics[width=75mm]{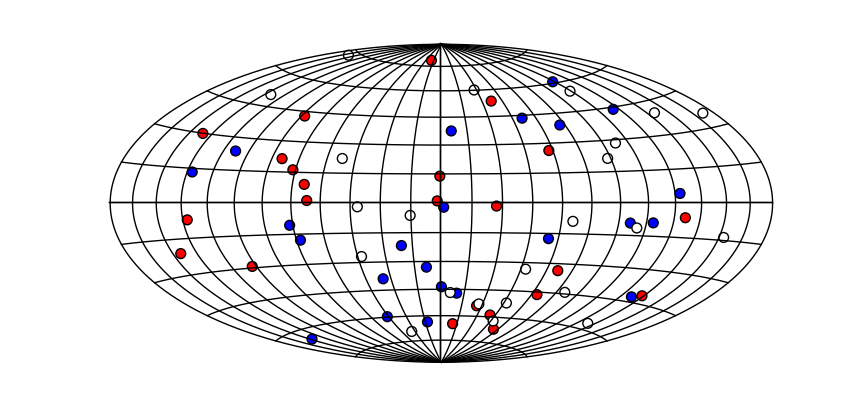}&\includegraphics[width=75mm]{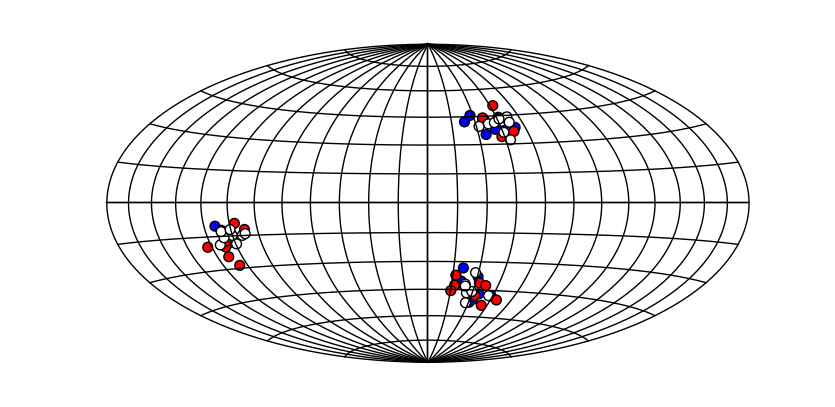}&\includegraphics[width=75mm]{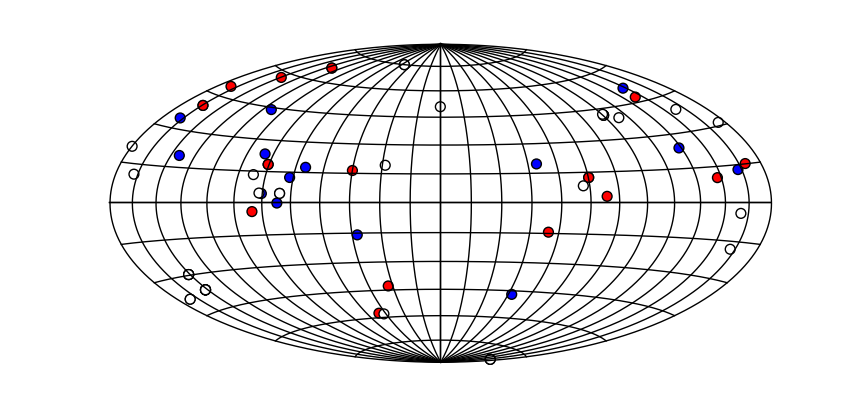}\\

\, & \includegraphics[width=75mm]{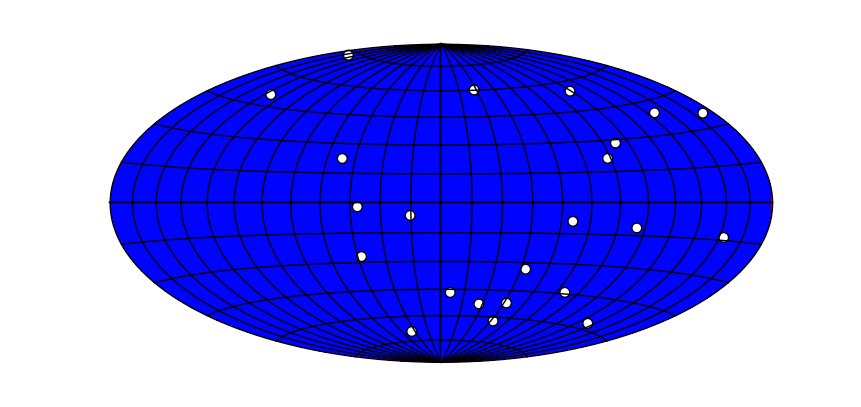}&\includegraphics[width=75mm]{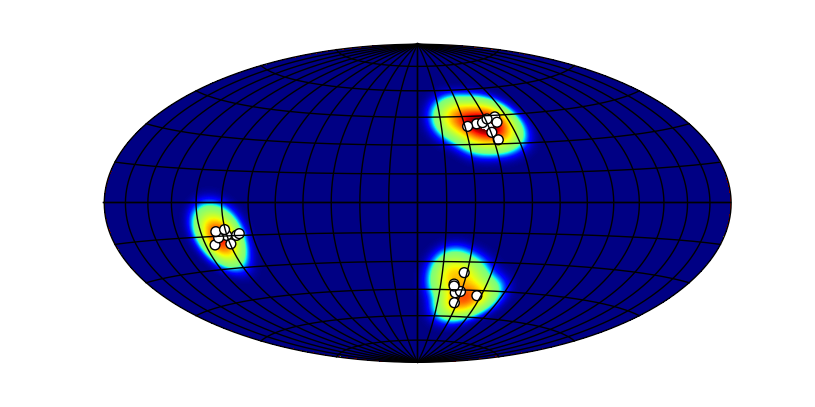}&\includegraphics[width=75mm]{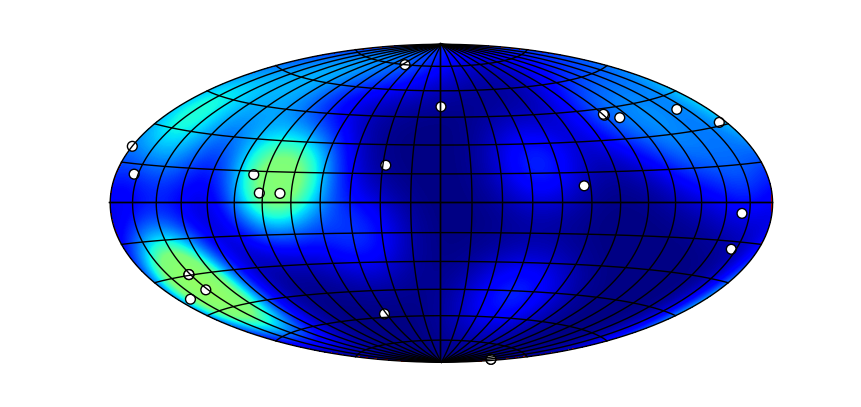}\\
\, & \includegraphics[width=75mm]{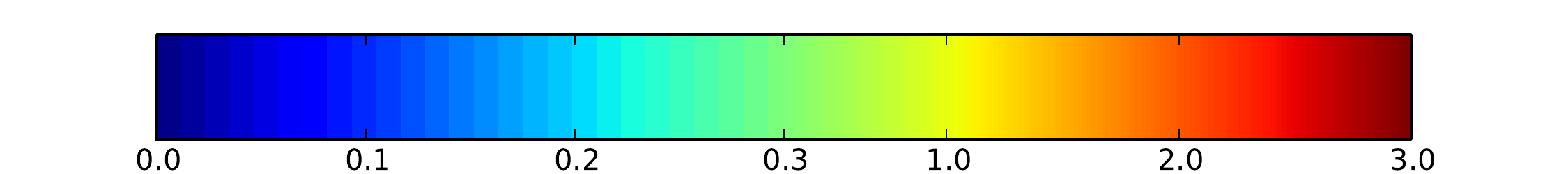}&\includegraphics[width=75mm]{Figures/colocolo.pdf}&\includegraphics[width=75mm]{Figures/colocolo.pdf}
\end{array}$
\end{center}

\caption{ Full process of model creation for data sets of 69 UHECRs for three test cases: uniform arrival directions (left); three sources (middle); and AGN sources from a realistic mock catalogue (right). Three aspects of the analysis procedure are shown: (1) the full input UHECR data set; (2) partition of the data set into  $\,\bluedot\,$  generating points,  $\,\reddot\,$  fitting points and  $\,\whitedot\,$  testing points; and (3) the resultant mixture distribution of vMF kernels centred on the generating points. The model is created for a range of $\kappa$ values, but for each of the three test cases, only the maximum likelihood value of $\kappa$ is displayed here. These highest likelihood values are $\kappa=\,$0, 108 and 13 for the uniform sources, three sources, and AGN sources, respectively. }
\label{fig:TriModel}
\end{figure}

\end{landscape}
\clearpage

\subsection{Non-uniform model}
\label{sec:nonuniform}
In contrast to the above uniform model, there is an infinite variety of possible non-uniform models that might explain the distribution of UHECR arrival directions. This is a significant conceptual problem: it is difficult to decide which alternative clustered model should be used. To resolve this issue, we develop a multi-stage, Bayesian approach by splitting the arrival directions $\{\bmath{r}_i\}$ into three subsets:
\begin{enumerate}
\item
First, $N_{\rm g}$ generating points $\{\bmath{r}_g\}$ are chosen as the centres of smooth, localized kernels which can be combined into a mixture distribution on the sphere (Section~\ref{sec:DerivModel}).
\item 
Then, $N_{\rm f}$ fitting points $\{\bmath{r}_f\}$ (with $f \in \{1, 2, \ldots, N_{\rm f}\}$) are used to obtain a distribution for the unspecified width parameter of the kernels (Section~\ref{sec:priordist}).
\item
Finally, the remaining $N_{\rm t}$ testing points $\{\bmath{r}_t\}$ (with $t \in \{1, 2, \ldots, N_{\rm t}\}$) are used to evaluate the marginal likelihood under this non-uniform model (Section~\ref{sec:ProbGivNew}).
\end{enumerate}
The partitions of the data are chosen at random and the generating points are not linked to the putative UHECR sources in any way.  This method is hence independent of any source catalogue or propagation model and, indeed, could be applied to any sample of points on the sphere.
The three steps of this approach are illustrated in Figure~\ref{fig:TriModel} for the three test cases described in Section~\ref{sec:TestCases}. 

The resultant model (and marginal likelihood) is fully specified, but the algorithm for generating it has two free parameters: $N_{\rm g}$ and $N_{\rm f}$.  The relative merits of using a low or high fraction of the data to generate and fit the model (leaving, respectively, a high or low fraction to evaluate the marginal likelihood) is an important area of investigation (e.g.\ \citealt{Spiegelhalter1982,OHagan1995}) but here we take the simplest approach by using a third of the data at each step, so $N_{\rm f} = N_{\rm g} = {\rm floor} (N / 3)$, leaving $N_{\rm t} = N - (N_{\rm f} + N_{\rm g}) \simeq N / 3$ testing points. The results of varying these divisions are deferred to a later paper.

The above three-step approach is novel, but similar in principle to the methods of partial or incomplete Bayes factors that have been explored by e.g.\ \cite{Spiegelhalter1982}, \cite{Aitkin1991}, \cite{OHagan1991}, \cite{OHagan1995} and \cite{Ghoshetal2006}. In all cases the aim is to evaluate the marginal likelihood for a model with unspecified parameters that do not have strongly motivated priors; and in all cases the basis of the approach is the same as is used here, namely to use part of the data to generate the parameter distributions that are necessary to evaluate the integral in Equation~\ref{eq:Evidence}. 

\subsubsection{ Generating a clustered model from the data }
\label{sec:DerivModel}

The first step to specifying a non-uniform model is to use the $N_{\rm g}$ generating points $\{\bmath{r}_g\}$ as the centres of smooth, localized kernels of an as yet unspecified angular size. 

The specific kernel chosen was the von Mises Fisher (vMF) distribution, which resembles a Gaussian on the sphere and is defined by the density
\begin{equation}
    \label{eq:smtng}
{\rm Pr}(\bmath{r} |\overline{\bmath{r}}, \kappa)= \frac{\kappa}{4\rmn{\pi}\sinh(\kappa)} e^{\kappa\bf{\bmath{r}}\cdot \overline{\bmath{r}}},
\end{equation}
where $\overline{\bmath{r}}$ is the central direction and $\kappa$ is the concentration parameter. This is inversely related to the width of the distribution: for large values of $\kappa$ the distribution is peaked over an angular scale of $\sim 1 / \sqrt{\kappa}$ , while if $\kappa$ tends to $0$ the distribution becomes uniform on the sphere.
The vMF distributions were centred on the generating points to give the mixture model density
\begin{equation}
\label{eq:mixture}
{\rm Pr}(\bmath{r} | \{\bmath{r}_g\}, \kappa)
=
 \frac{\kappa}{4\rmn{\pi}N_{\rm g}\sinh(\kappa)} \sum_{g=1}^{N_{\rm g}} e^{\kappa \bmath{r} \cdot \bmath{r}_g}.
\end{equation}

\subsubsection{Obtaining a concentration distribution}
\label{sec:priordist}

The last step to fully defining the non-uniform model is to specify a distribution for $\kappa$. This is done by using the fitting points to obtain a fully normalized posterior for $\kappa$ that can be used as a parameter prior in the model comparison step. A uniform prior for $\kappa \geq 0$ is chosen in order to include models with $\kappa=0$ (which would not be possible for, e.g. a logarithmic prior in $\kappa$). The posterior distribution that results from generating points $\{\bmath{r}_g\}$ and fitting points $\{\bmath{r}_f\}$ is
\begin{eqnarray}
{\rm Pr}(\kappa|\left\{\bmath{r}_{g}\right\},\left\{\bmath{r}_{f}\right\}) & = & \frac{{\rm Pr}(\kappa ) \,{\rm Pr}(\left\{\bmath{r}_{f}\right\}|\left\{ \bmath{r}_{g}\right\},\kappa)}{\int_{0}^{\infty}{\rm Pr}(\kappa ' )\,{\rm Pr}(\left\{\bmath{r}_{f}\right\}|\left\{ \bmath{r}_{g}\right\},\kappa ')\,{\rm d}\kappa '}
\nonumber \\
\label{eq:KappaPost2}
& \propto & \Theta(\kappa) \prod_{f=1}^{N_{\rm f}} {\rm Pr}(\bmath{r}_f | \{ \bmath{r}_g \}, \kappa) \nonumber \\
& \propto & \frac{\Theta(\kappa) \, \kappa^{N_{\rm f}}}{\sinh^{N_{\rm f}}(\kappa)} \prod_{f=1}^{N_{\rm f}} \left( \sum_{g = 1}^{N_{\rm g}} e^{\kappa \bmath{r}_f \cdot \bmath{r}_g} \right),
\end{eqnarray}
where $\Theta(\kappa)$ is the Heaviside step function that encodes the fact that $\kappa$ is non-negative.  The posterior distribution is straightforward to normalize numerically as it is (generally) unimodal and as there is only one parameter. 

The alternative, non-uniform model for the UHECR arrival directions is hence fully specified (in the sense of being usable in Bayesian model comparison).  It is a sum of vMF distributions centred on the set of generating points, $\{\bmath{r}_g\}$, and with the distribution of vMF concentration parameter $\kappa$ given by Equation~\ref{eq:KappaPost2}.   

\subsubsection{Evaluating the marginal likelihood}
\label{sec:ProbGivNew}
Having specified the non-uniform model, $M_{\rm n}$, with the generating points, $\{ \bmath{r}_g\}$ and obtained the distribution ${\rm Pr}(\kappa|M_{\rm n})$ by using the fitting points, $\{ \bmath{r}_f\}$, it is now possible to use the remaining data, the testing points $\{ \bmath{r}_t\}$, to evaluate the marginal likelihood.  From Equation~\ref{eq:Evidence} this is 
\begin{equation}
\label{eq:ProbGen}
{\rm Pr}( \left\{ \bmath{r}_{t} \right\} | M_{\rm n} ) 
 =  \int_0^\infty {\rm Pr}(\kappa | M_{\rm n}) \, {\rm Pr}( \left\{ \bmath{r}_{t} \right\} | \kappa,  M_{\rm n}) \, {\rm d} \kappa,
\end{equation}
where ${\rm Pr}(\kappa | M_{\rm n}) = {\rm Pr}(\kappa | \{ \bmath{r}_{g} \}, \left\{ \bmath{r}_{f} \right\})$ is given in Equation~\ref{eq:KappaPost2} and now plays the role of the prior distribution for $\kappa$, and the likelihood for the testing points is (cf.\ Equation~\ref{eq:mixture})
{\setlength
\arraycolsep{1pt}
\begin{eqnarray}
{\rm Pr}(\{\bmath{r}_{t} \} | \kappa , M_{\rm n}) 
&=&
{\rm Pr}(\{\bmath{r}_{t} \} | \{\bmath{r}_{g} \}, \kappa ) 
\nonumber \\
&=&
\prod_{t = 1}^{N_{\rm t}} {\rm Pr}(\bmath{r}_t | \{ \bmath{r}_g\}, \kappa) \nonumber \\
&=&
\frac{\kappa^{N_{\rm t}}}{[4\rmn{\pi}N_{\rm g}\sinh(\kappa)]^{N_{\rm t}}} \prod_{t=1}^{N_{\rm t}} \left( \sum_{g = 1}^{N_{\rm g}} e^{\kappa \bmath{r}_t \cdot \bmath{r}_g} \right). 
\end{eqnarray}}
The one-dimensional integral in Equation~\ref{eq:ProbGen} is, once again, straightforward to evaluate numerically. This then gets further modified by the non-uniform exposure, as described in Section~\ref{sec:exposure}.

\begin{figure*}
  \centering $
\arraycolsep=0.01pt\def\arraystretch{-0.5}
\begin{array}{ccc}
\includegraphics[width=90mm]{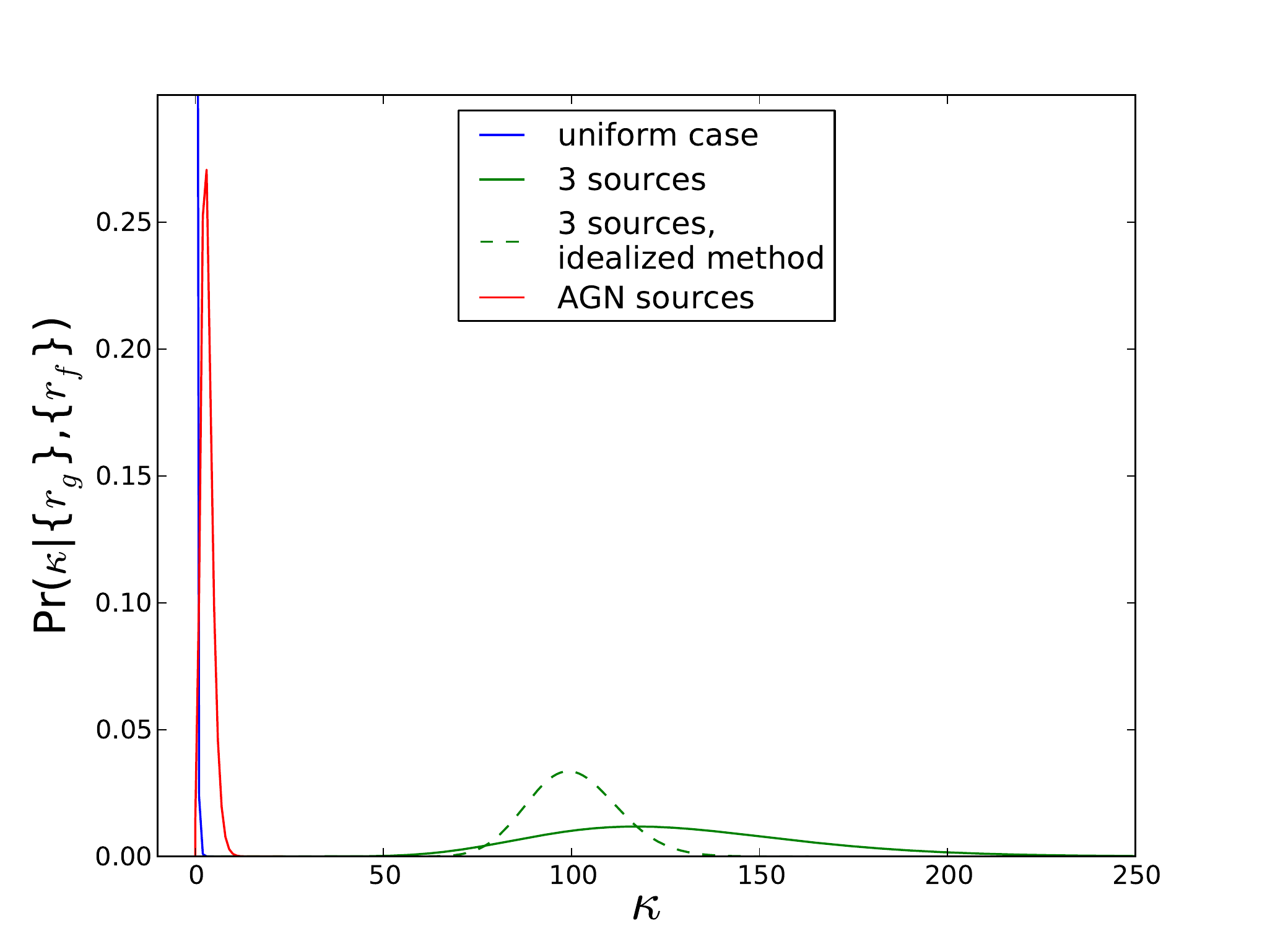}&
\includegraphics[width=90mm]{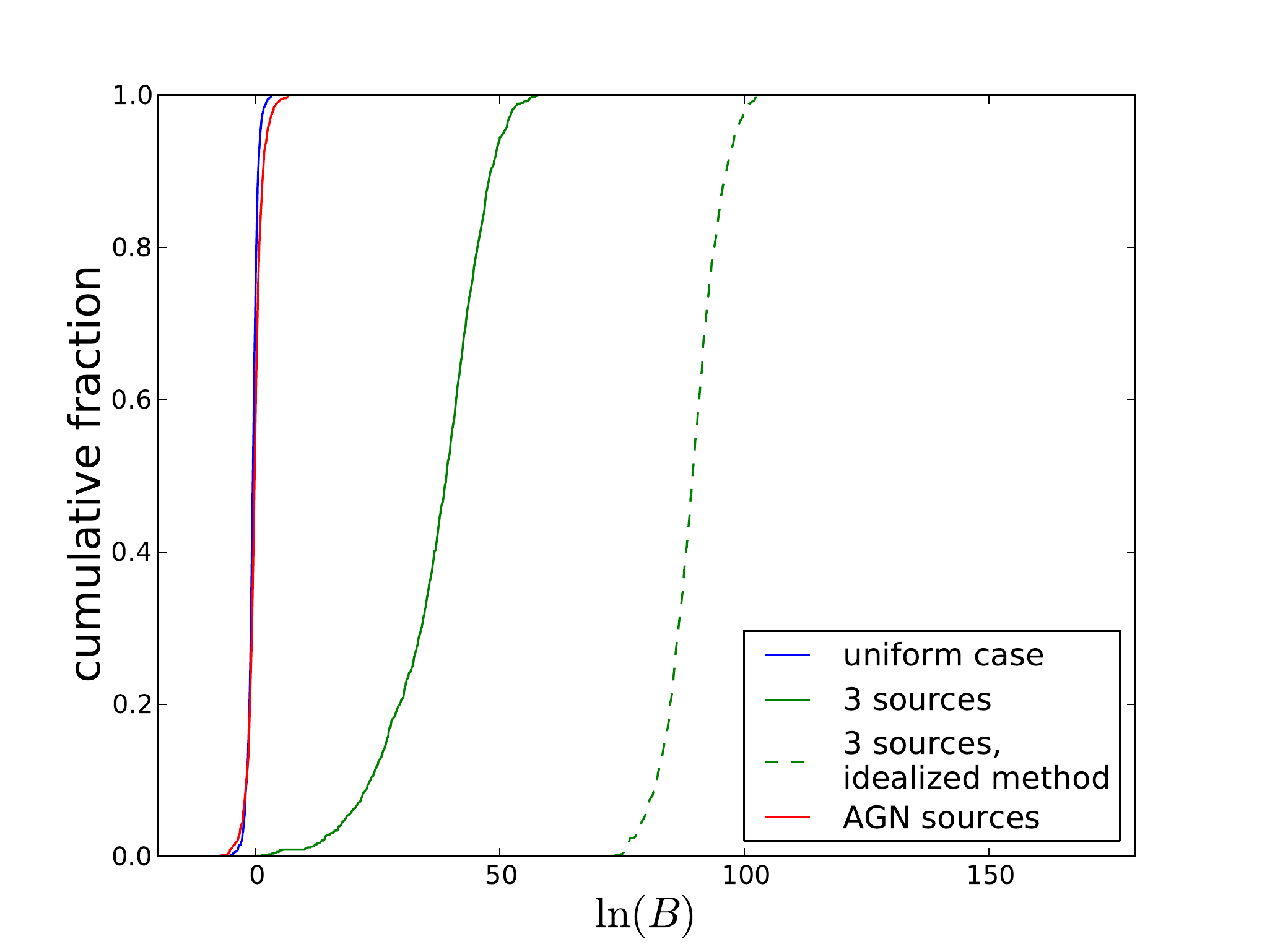}
\end{array}$
  \caption{(A) Kappa posteriors and (B) cumulative fractions of Bayes factors, produced by the application of the multi-step method to test cases of 69 UHECR events. Three test cases are considered: uniform UHECRs; UHECRs generated from three sources; and UHECRs generated by AGNs from a realistic catalogue. In the case of three sources, in addition to the conventional application of the multi-step method, the results for an idealized method are displayed. In the idealized application of the method, the generating points are taken as the true centres of the vMF distributions that generate the UHECRs, rather than as a random subset of the data.}
\label{fig:TwoThings}
\end{figure*}

\subsection{Non-uniform exposure }
\label{sec:exposure}

When studying the measured arrival directions of CRs in a real experiment, the non-uniform exposure of the observatory needs to be taken into account.  This is characterized by the relative exposure per unit solid angle, ${\rm d}\epsilon / {\rm d}\Omega$, defined such that $\int ({\rm d}\epsilon / {\rm d}\Omega)\, {\rm d}\Omega =\epsilon_{{\rm tot}}$ is the total exposure\footnote{The units of the total exposure are km$^2$ sr yr.}. The relative exposure is proportional to ${\rm  Pr}({\rm det}|\bmath{r})$, the probability that a UHECR arriving from direction $\bmath{r}$ is detected.  The distribution of arrival directions of detected CRs is then given by Bayes's theorem as
\begin{equation}
    \label{eq:Equation WithExposure}
{\rm Pr}( \bmath{r} | {\rm det} ) \propto {\rm Pr}(\bmath{r})\,{\rm Pr}({\rm det}|\bmath{r})\propto {\rm Pr}(\bmath{r})\,\frac{{\rm d}\epsilon}{{\rm d}\Omega},
\end{equation}
where ${\rm Pr}(\bmath{r})$ is the distribution of arrival directions of all CRs, irrespective of whether they are detected.

For uniform UHECR arrival directions discussed in Section~\ref{sec:uniform}, $ {\rm Pr}(\bmath{r}|M_{\rm u})=1/(4\pi)$, so that  ${\rm Pr}( \bmath{r}, {\rm det} | E )$ simply becomes 
\begin{equation}
    \label{eq:Equation WithExposure}
{\rm Pr}( \bmath{r} | {\rm det},M_{\rm u} ) = \frac{1}{\epsilon_{{\rm tot}}}\frac{{\rm d}\epsilon}{{\rm d}\Omega} .
\end{equation}

For the non-uniform UHECR arrival directions discussed in Section~\ref{sec:nonuniform}, $ {\rm Pr}(\bmath{r}|\kappa,M_{\rm n})$ is given in Equation~\ref{eq:mixture}, so that
\begin{equation}
    \label{eq:smtng_mod}
{\rm Pr}( \bmath{r}| {\rm det} ,\kappa , M_{\rm n} ) \propto  \frac{{\rm d}\epsilon}{{\rm d}\Omega} \sum_{g = 1}^{N_{\rm g}} e^{\kappa \bmath{r} \cdot \bmath{r}_g},
\end{equation}
where the normalization depends on the position of the generating points, $\{\bmath{r}_g\}$, the relative exposure and $\kappa$, and must be calculated numerically.

\subsection{Illustration of the multi-step method}
\label{sec:TestCases}

Figure~\ref{fig:TriModel} illustrates how the multi-step Bayesian method works for several simple test cases: a uniform source distribution; a model with three sources; and a model based on the AGN simulations described below in Section~\ref{sec:AGN}. The total number of UHECRs is 69 in all cases. The associated $\kappa$ posteriors and the resultant distribution of Bayes factors are shown in Figure~\ref{fig:TwoThings}. 

The first test case was a very simple scenario: the UHECRs were simulated with isotropic arrival directions, for the case of uniform exposure. The $\kappa$ posterior for the uniform case has its maximum very close to 0, and declines rapidly, because the vMF distributions that are fitted to the data are almost uniform. The Bayes factors for this case are small: the uniform model is favoured in 74.1\% of the simulations. 

The second test case is a simple model of non-uniform arrival directions: the UHECRs were sampled from three vMF distributions, representing three UHECR sources. The concentration parameter $\kappa$ of the vMF distributions was taken as 90. The $\kappa$ posterior for this case is systematically peaked at higher values, as can be seen in Figure~\ref{fig:TwoThings}A. It is peaked at a value higher than the input value of $\kappa$ because each of the three original kernels is now accounted for by multiple narrower kernels that are slightly off-centre. The Bayes factors are very large: the non-uniform model is favored in more than 99.9\% of the simulations and the average Bayes factor is $\sim 40$.

For the case of three sources, it was also possible to apply an idealized form of the multi-step method: instead of using one third of the full data set as the generating points, the generating points were taken as the actual positions of the sources of the UHECRs. In this way, the idealized method does not share the catalogue-indepence of the full three-step method described in Section~\ref{sec:nonuniform}. For this idealized case, the $\kappa$ posterior is consistent with the input value, because the three original kernels are accounted for by three kernels located on the original kernel positions. This is also the reason why the Bayes factors are even larger than for the ordinary case.
The idealized form of the multi-step method is useful to see the potentially strong impact the lack of knowledge about the source positions can have, although it hence cannot be used to analyze real data.

The third test case was the case for UHECRs generated by AGNs, simulated with the realistic model described in Section~\ref{sec:AGN}. The input value of $\kappa=360$ was chosen to give the strongest plausible signal, but the resultant posterior is peaked close to $\kappa=0$. The reason is that there are now so many sources compared to the number of UHECRs that the source distribution is undersampled. This is an indication that, given the weak (projected) clustering expected of nearby AGNs, a significantly larger UHECR sample would be needed for their self-clustering to be apparent. More realistic tests that are documented in Section~\ref{sec:ApplicationSimul} confirm this result.
 
\section{Application to simulated UHECR samples}
\label{sec:ApplicationSimul}

To investigate the effectiveness of the multi-stage Bayesian method described above, it was applied to realistic mock catalogues of UHECRs. Catalogues were created for two different UHECR scenarios: isotropic (Section~\ref{sec:Iso}) and AGN centred (Section~\ref{sec:AGN}). The samples of incoming UHECRs were then subjected to the PAO measurement process (Section~\ref{sec:Meas}). The distributions of Bayes factors for the resultant observed samples are analysed in Section~\ref{sec:SimResults}.

\subsection{Isotropic distribution of sources}
\label{sec:Iso}

The application of the multi-step method to uniform UHECR distributions acted as a false positive test. Computing large numbers of Bayes factors for uniform UHECR distributions can be used to establish how often the null hypothesis is wrongly rejected.

\subsection{AGN sources}
\label{sec:AGN}

Simulated UHECR catalogues were created for the case of UHECRs originating in AGNs. The simulation encompassed two main components: the injection of the UHECRs at the sources and a propagation model.

\subsubsection{Injection at the sources}

The AGN sources were drawn randomly from the simulated Las Damas ``Consuelo" catalogues\footnote{http://lss.phy.vanderbilt.edu/lasdamas/}, following a similar procedure to \cite{Berlind2011}.

Two source densities were used: $10^{-3.5}\,{\rm Mpc}^{-3}$ and $10^{-4.5}\,{\rm Mpc}^{-3}$. These are the highest and lowest source densities available in the Consuelo catalogues, and represent a reasonable range of possible source densities.

The injection spectrum of the UHECRs at the sources is assumed to be a power-law of the form $Q(E)\propto E^{-\alpha}$, where $Q(E)\,{\rm d}E$ is the number of cosmic rays emitted with energy between $E$ and $E + {\rm d}E$ per unit time, and $\alpha$ is the power law index. Simulations were conducted for three realistic values of the index: 2.0, 2.3 and 2.7, spanning the range of values used in e.g.\ \cite{Domenico2013}, \cite{AugerBounds2013}, \cite{Ahlers2011} and \cite{Decerprit2012}.

\subsubsection{Propagation model}

Both the energy loss that the UHECRs experience during propagation and their magnetic deflection must be accounted for. The deflection is not treated explicitly, but included in the observational smearing described in Section~\ref{sec:Meas}; the energy loss model is described here.

A pure proton composition of UHECRs was assumed and so the energy loss during propagation consists of three components (e.g.\ \citealt{Stanev2009}):
\begin{enumerate}
\item
The GZK scattering off the CMB photons at energies above $E \ga 5\times 10^{19}\, {\rm eV}$;
\item
Bethe Heitler ${\rm e^{+}e^{-}}$ pair production (also a scattering process off the CMB radiation), which dominates at lower energies (\citealt{Hillas1968});
\item
The adiabatic energy loss due to the expansion of the Universe.
\end{enumerate}
Our implementation of this propagation model includes the BH and adiabatic losses in a continuous approximation, and treats the GZK effect as a stochastic process.

\subsection{Measurement}
\label{sec:Meas}

All of the simulations were done for a PAO-like experiment, three aspects of which were modelled explicitly:
\begin{enumerate}
\item
PAO's non-uniform exposure was taken into account by accepting arriving UHECRs with a probability proportional to the relative exposure ${\rm d}\epsilon/{\rm d}\Omega$ defined in Section~\ref{sec:exposure}.
\item
The error in PAO's energy measurement is about $12\%$ (\citealt{2013arXiv1310.4620L}), and was included in the model. This is significant as only UHECRs that have an observed energy above a fixed threshold are included in the simulated samples.
\item
The angular resolution of PAO varies from about $2.2$ deg to about $1$ deg for the lowest and highest energies respectively (\citealt{PAO2012}). The magnetic deflection that the UHECRs experience during propagation also means that their arrival directions are offset from the source. The magnitude of this effect is uncertain, the estimates of typical deflection angles ranging from  $\sim 2$ to  $\sim 10$ deg for the highest energy UHECRs (e.g. \citealt{Medina1998,Sigl2004,Dolag2005}). These two effects are simulated together by drawing a measured arrival direction from a vMF distribution centred on the source. We used three different values for the concentration parameter $\kappa$ of the vMF distributions: 30, 90, and 360, which correspond to average angular  deviations of approximately 10, 6 and 3 deg, respectively.\par
We treated the magnetic deflection as a simple smearing, rather than including detailed simulations of the Galactic and extra-Galactic magnetic fields, because our aim was to assess the arrival directions without reference to a particular physical model. Detailed models of the magnetic fields are available (\citealt{Domenico2013,Farrar2014} and references therein), and a formalism for incorporating these into a Bayesian UHECR analysis has been developed in \cite{Soiaporn2012}.

\end{enumerate}

\subsection{Results of the simulations}
\label{sec:SimResults}

Simulations were performed and Bayes factors  evaluated for the isotropic model, and for the AGN-centred model with 18 combinations of the above parameters:
\begin{enumerate}
\item
source densities of $10^{-3.5}\,{\rm Mpc}^{-3}$ and $10^{-4.5}\,{\rm Mpc}^{-3}$;
\item
injection parameters $\alpha$ of $2.0$, $2.3$, $2.7$;
\item
concentration parameters $\kappa$ of  $30$, $90$, $360$. 
\end{enumerate}
For each of the 18 combinations of parameters, 1,000 samples of 69 UHECRs were created (matching the size of the PAO data set). For each sample, Bayes factors were computed for each of three energy thresholds: $5.5 \times 10^{19}\, {\rm eV}$, $8.0 \times 10^{19}\, {\rm eV}$, and $10 \times 10^{19}\, {\rm eV}$. Including the 1,000 realisations of the isotropic model, 55,000 Bayes factors were computed in total.

\begin{figure}
\begin{center}
\hbox{\hspace{23ex}\includegraphics[width=120mm]{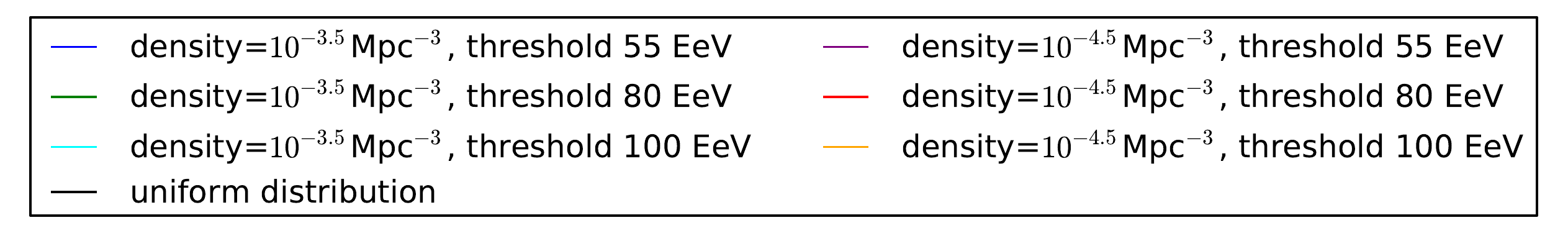}}
\end{center}
\hspace{5cm}
\vspace{-1cm}
\caption{Results of the multi-step method applied to mock UHECR catalogues. Cumulative distributions of Bayes factors have been produced for three energy thresholds, two source densities, and for different values of the sample size $N$, the injection parameter $\alpha$ and the concentration parameter $\kappa$, as indicated above.} 
\label{fig:AllOfThem}
\end{figure}

The results of these simulations are shown as cumulative distributions of Bayes factors in the top half of Figure~\ref{fig:AllOfThem}. These are compared to similar cumulative distributions for the case of uniformly distributed UHECRs.

The Bayes factors tend to be larger for the source-centred case than for the uniform case. The difference between the results for uniform and non-uniform UHECRs is greater for the case of low source density, as for higher source density the UHECR distribution would eventually tend to a uniform distribution.

Furthest away from the uniform case is the model with the lowest source density, highest $\kappa$ and highest $\alpha$. Higher $\kappa$ means that the UHECR arrival directions are more closely correlated with the positions of the sources. High $\alpha$ reduces the GZK horizon, meaning fewer contributing AGN sources and hence more non-uniformity.

The threshold energy value does not have a substantial effect on the distribution of Bayes factors. It is difficult to predict the effect of the threshold energy qualitatively, because there are two competing effects: a lower threshold would increase the sample size, which makes the non-uniformity more apparent; a higher threshold decreases the effective GZK horizon, which would increase the non-uniformity signature. This means that there is some ideal threshold that gives the greatest chance of detecting whatever anisotropy is present.

While the results for the uniform and non-uniform cases are clearly different, the difference is not very significant. If we take a threshold value of $\ln(B)=5$ to represent a decisive detection, then anisotropy is detected only for 0.002\% and 5\% of the samples for source densities of $10^{-3.5}\,{\rm Mpc}^{-3}$ and $10^{-4.5}\,{\rm Mpc}^{-3}$, respectively. The conclusion is that the clustering expected from a realistic model of AGN-sourced UHECRs is too weak to be detected from a sample of 69 events. This is consistent with the results of \cite{PAO2012}.

The simulations were repeated for 100 samples of $N = 690$ UHECRs (i.e.\ $10\,\times\,$the PAO sample). The results are shown in the bottom half of Figure~\ref{fig:AllOfThem}. The difference between the uniform and non-uniform cases becomes very apparent for all combinations of parameters. For source densities of $10^{-3.5}\,{\rm Mpc}^{-3}$ and $10^{-4.5}\,{\rm Mpc}^{-3}$, 22\% and 93\% of the Bayes factors are above the threshold of $\ln(B)=5$. UHECR samples of 690 events are sufficient to detect self-clustering for a realistic model.\par
We assume a pure proton composition of UHECRs, which is consistent with the results of HiRes (\citealt{Abbasi2005}), but not fully consistent with the results of PAO, which indicate a more complex mixed nuclear composition (\citealt{Unger2007}), including heavier nuclei such as iron. For iron, the magnetic deflection angle would be increased by a factor of 26, leading to a deflection of $\sim 50$ to  $\sim 250$ deg. This makes it more difficult to associate the UHECRs with specific sources. However, the detection of clustering is also made easier by the fact that heavier nuclei lose more energy through additional scattering processes, which reduces the GZK horizon and thus the number of candidate sources. The energy loss length for cosmic rays with $E\ga5\times 10^{20} {\rm eV}$ is reduced from $\sim 10 \,{\rm MeV}$ for protons to $\sim 2 \,{\rm MeV}$ for iron, which reduces the GZK horizon by a factor of $\sim$5 (\citealt{Stanev2009}). The net effect of these two factors will need to be established through additional simulations. 

\onecolumn
\clearpage

\begin{figure}
\begin{center}$
\arraycolsep=0.01pt\def\arraystretch{0.01}
\begin{array}{ccc}
\hbox{\includegraphics[width=40mm]{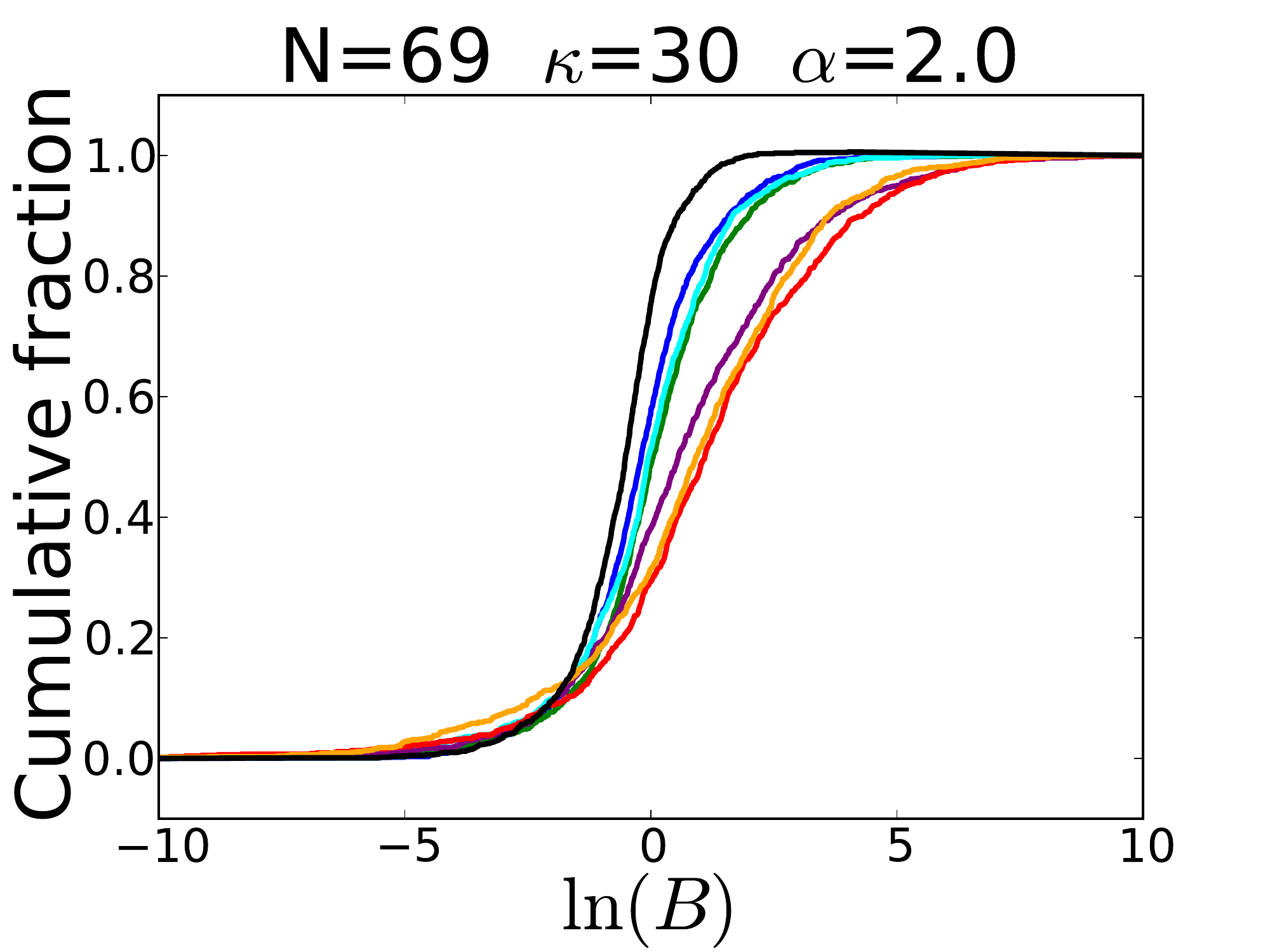}} & \includegraphics[width=40mm]{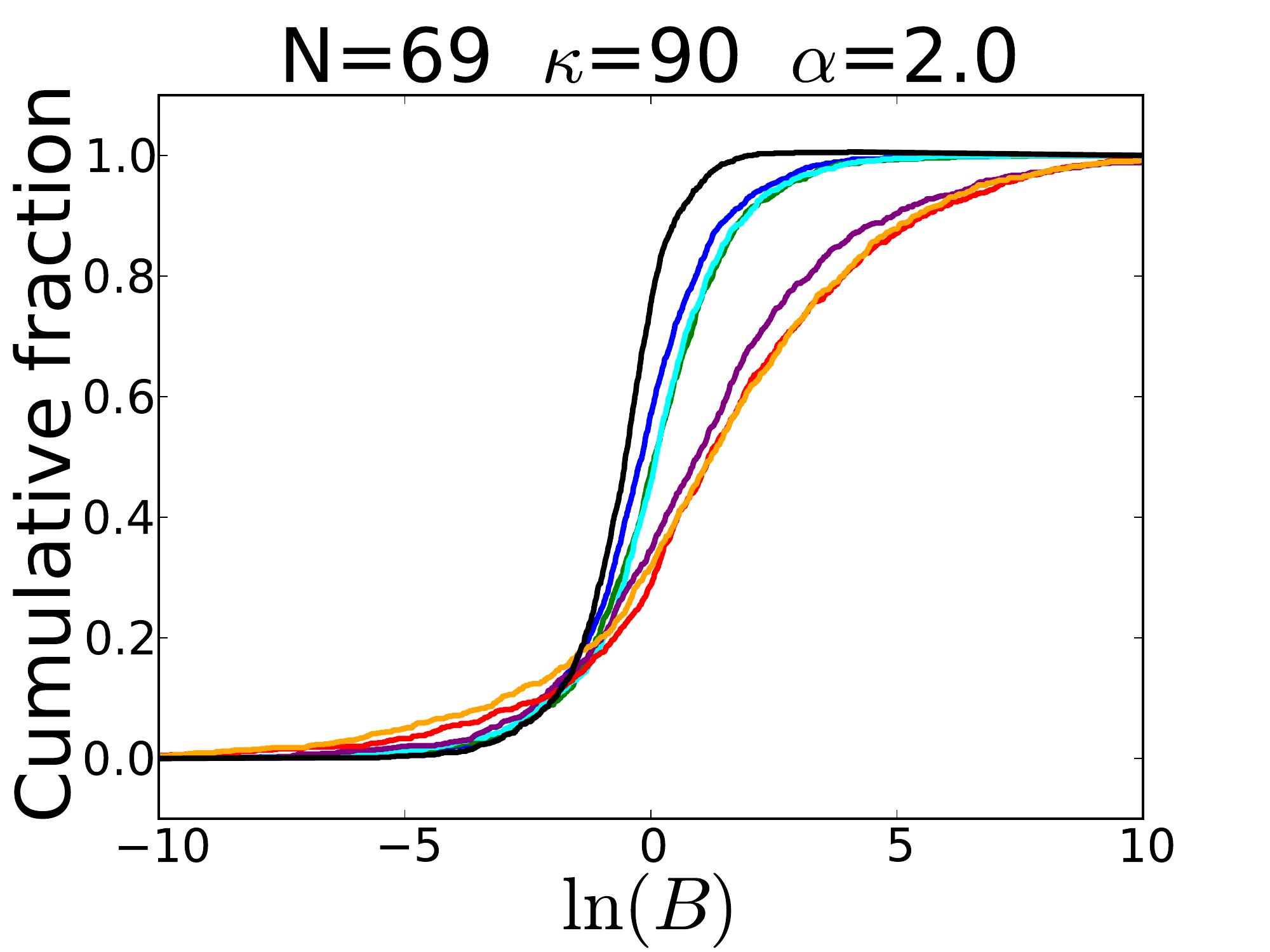} & \includegraphics[width=40mm]{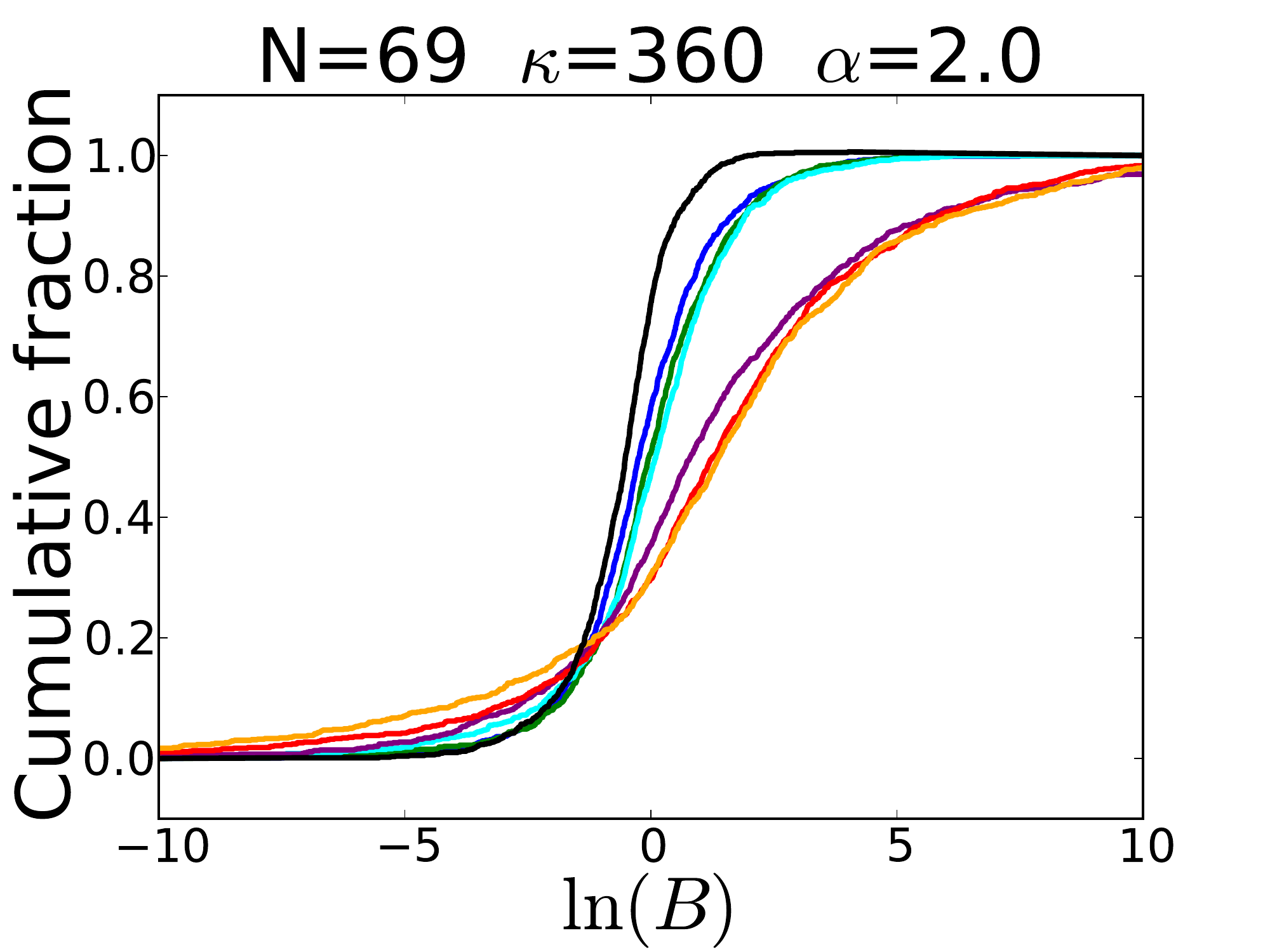} \\
\hbox{\includegraphics[width=40mm]{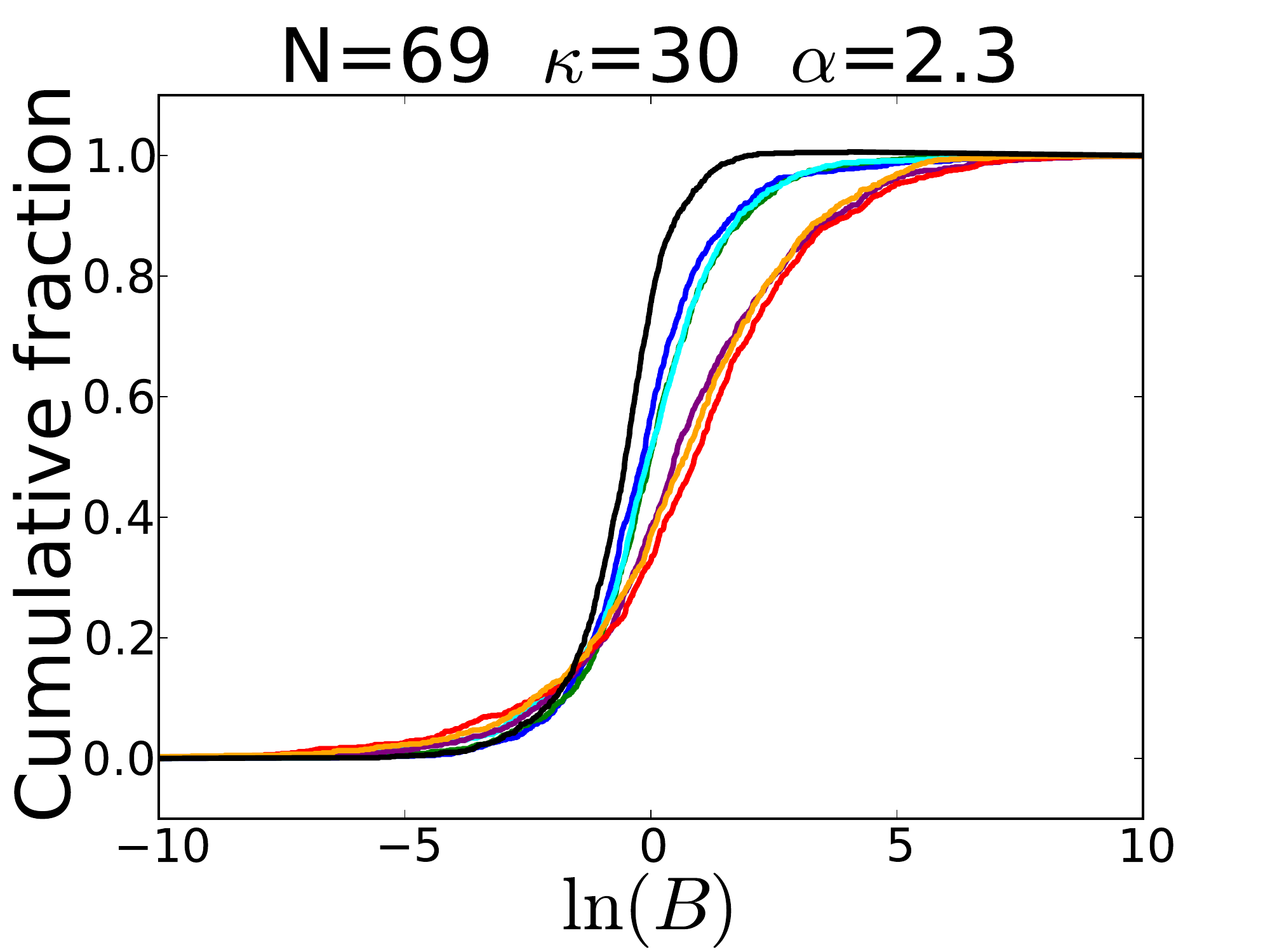}} & \includegraphics[width=40mm]{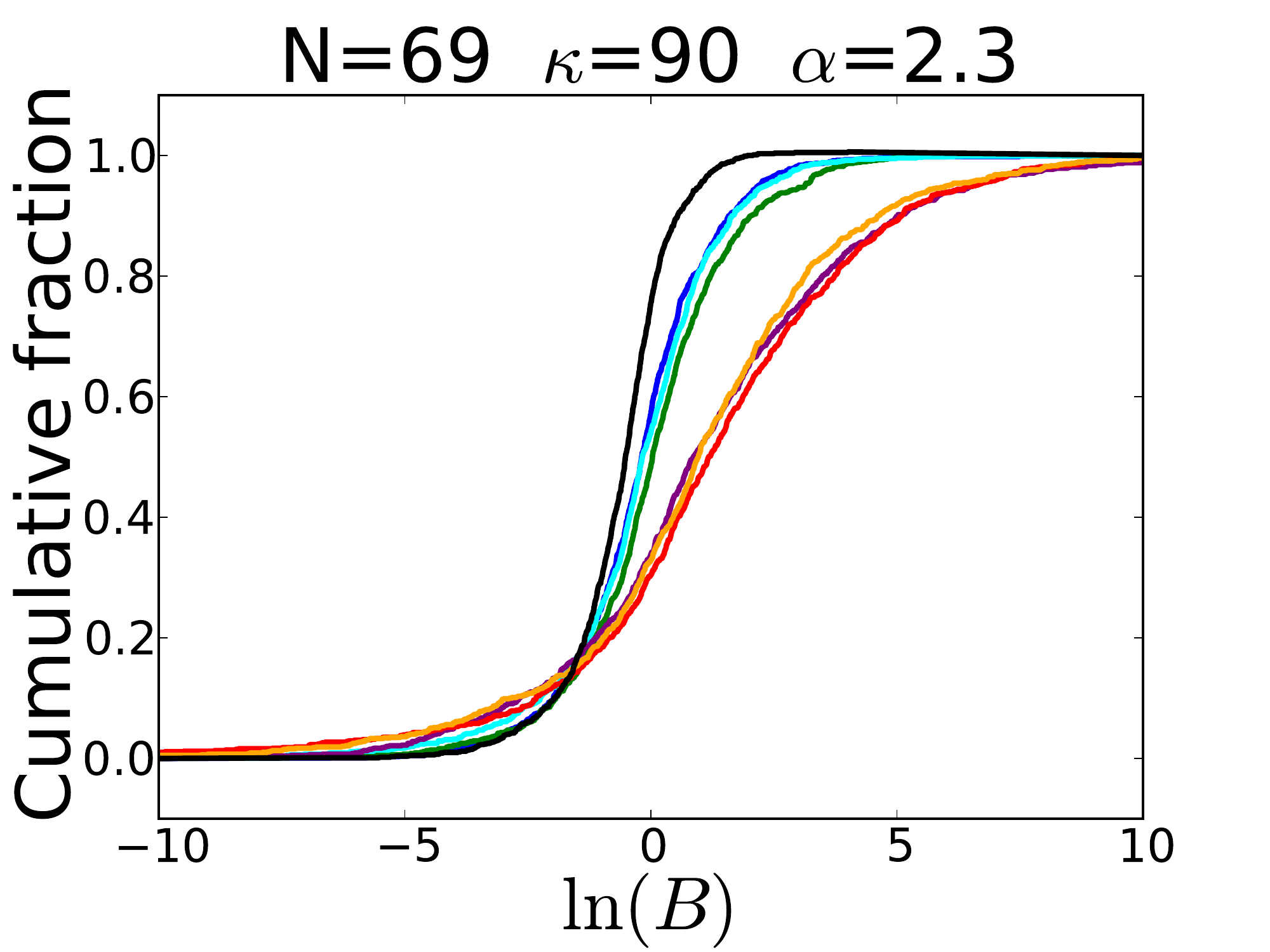} & \includegraphics[width=40mm]{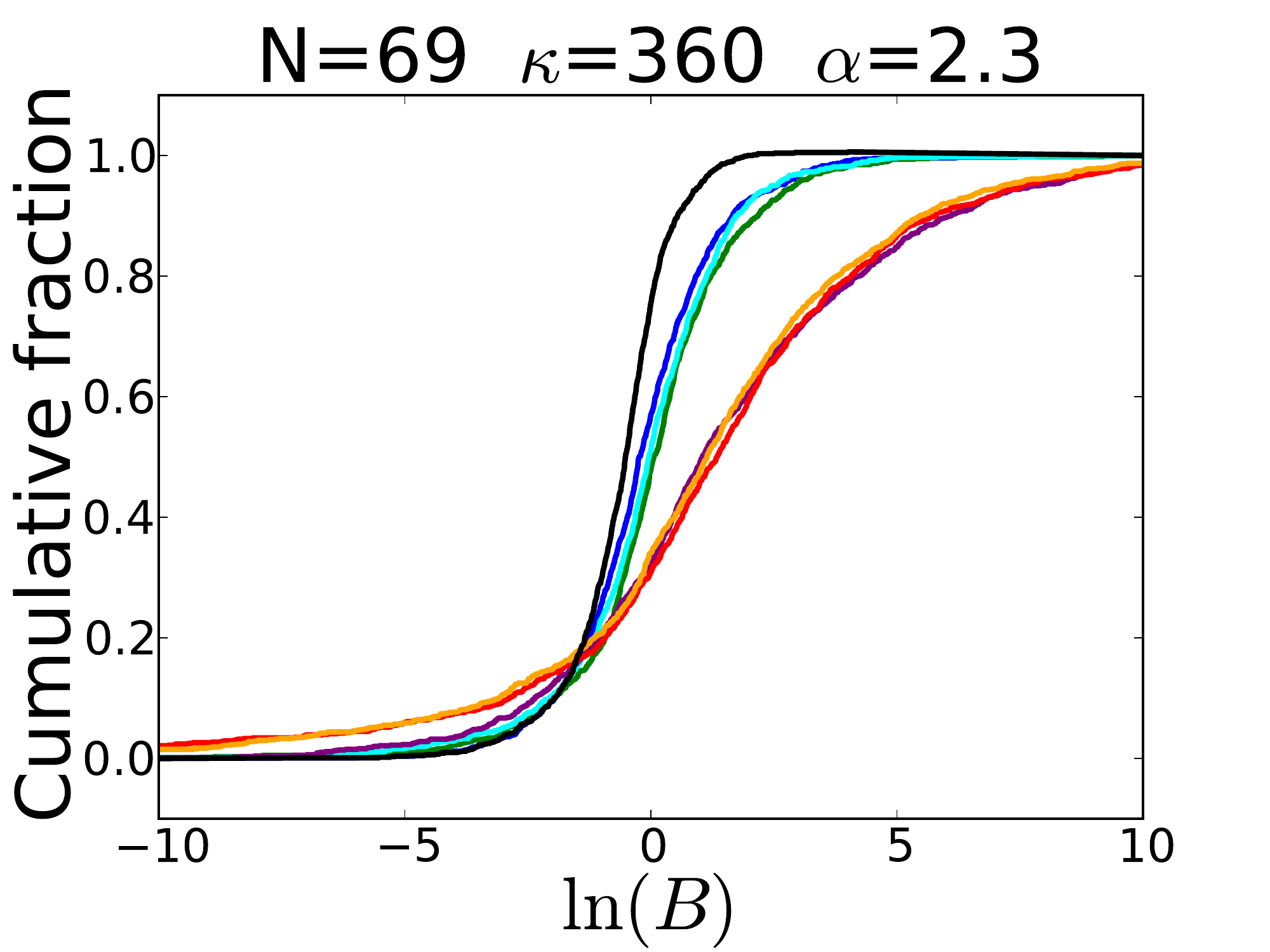} \\
\hbox{\includegraphics[width=40mm]{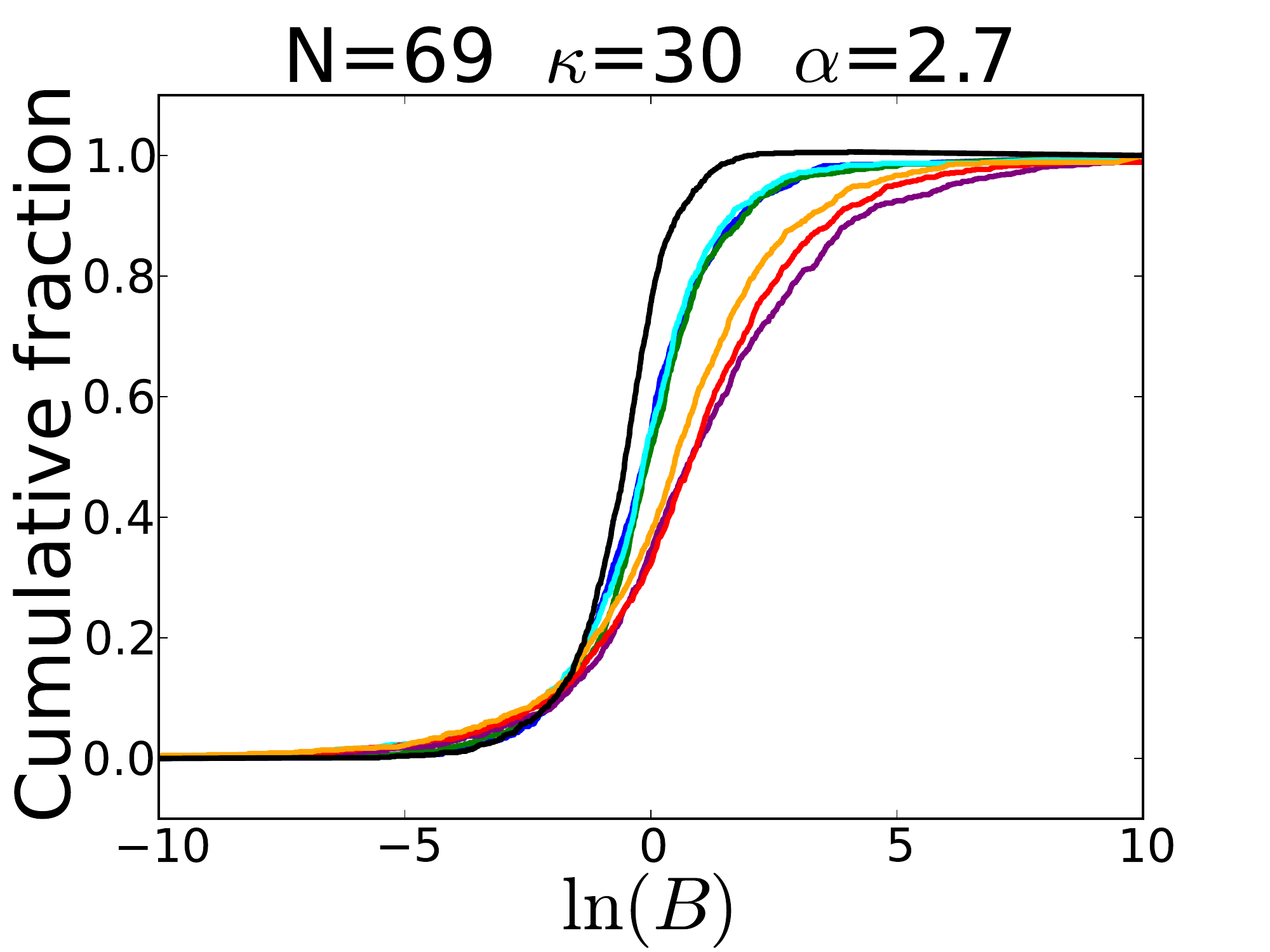}} & \includegraphics[width=40mm]{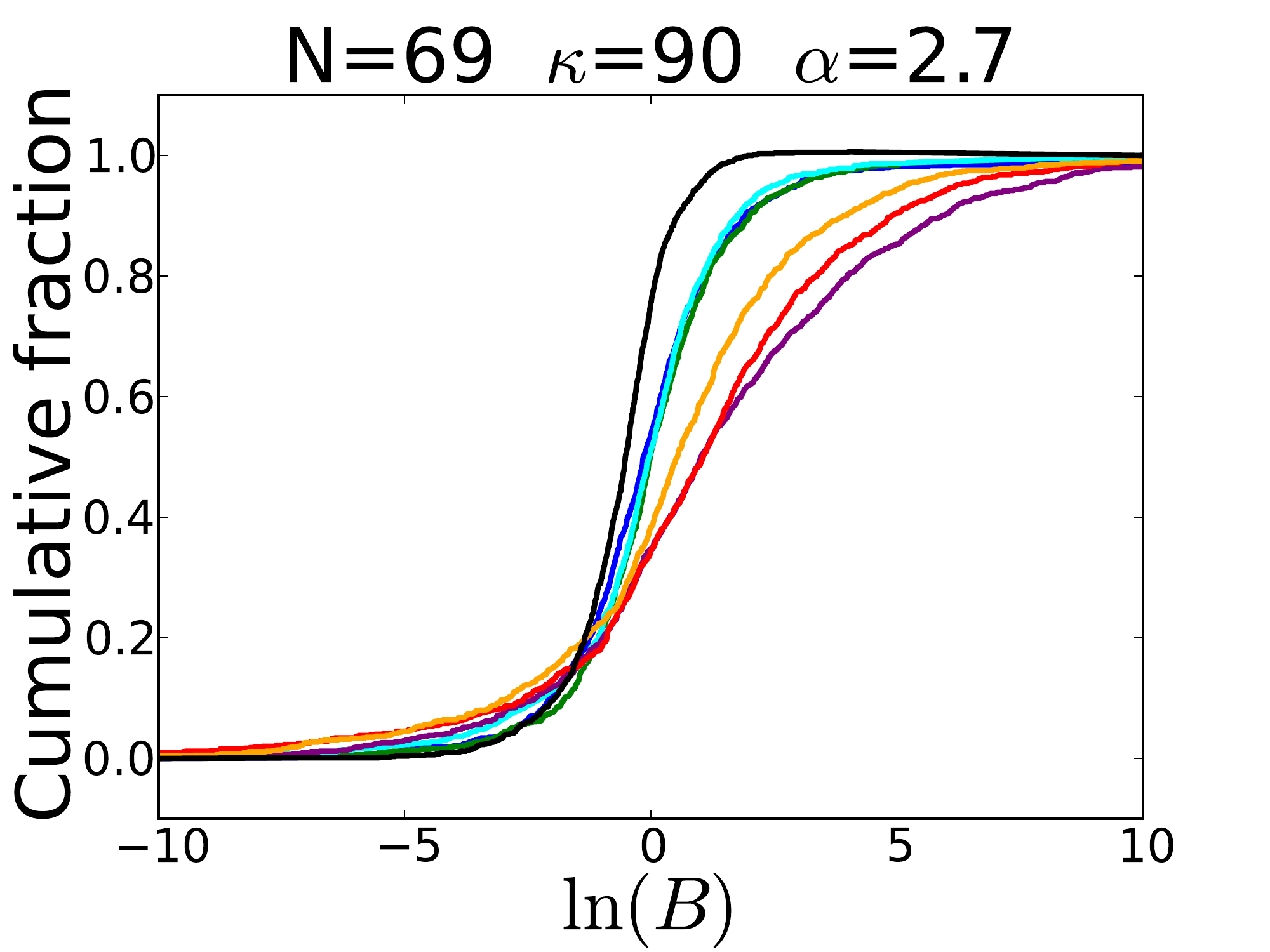} & \includegraphics[width=40mm]{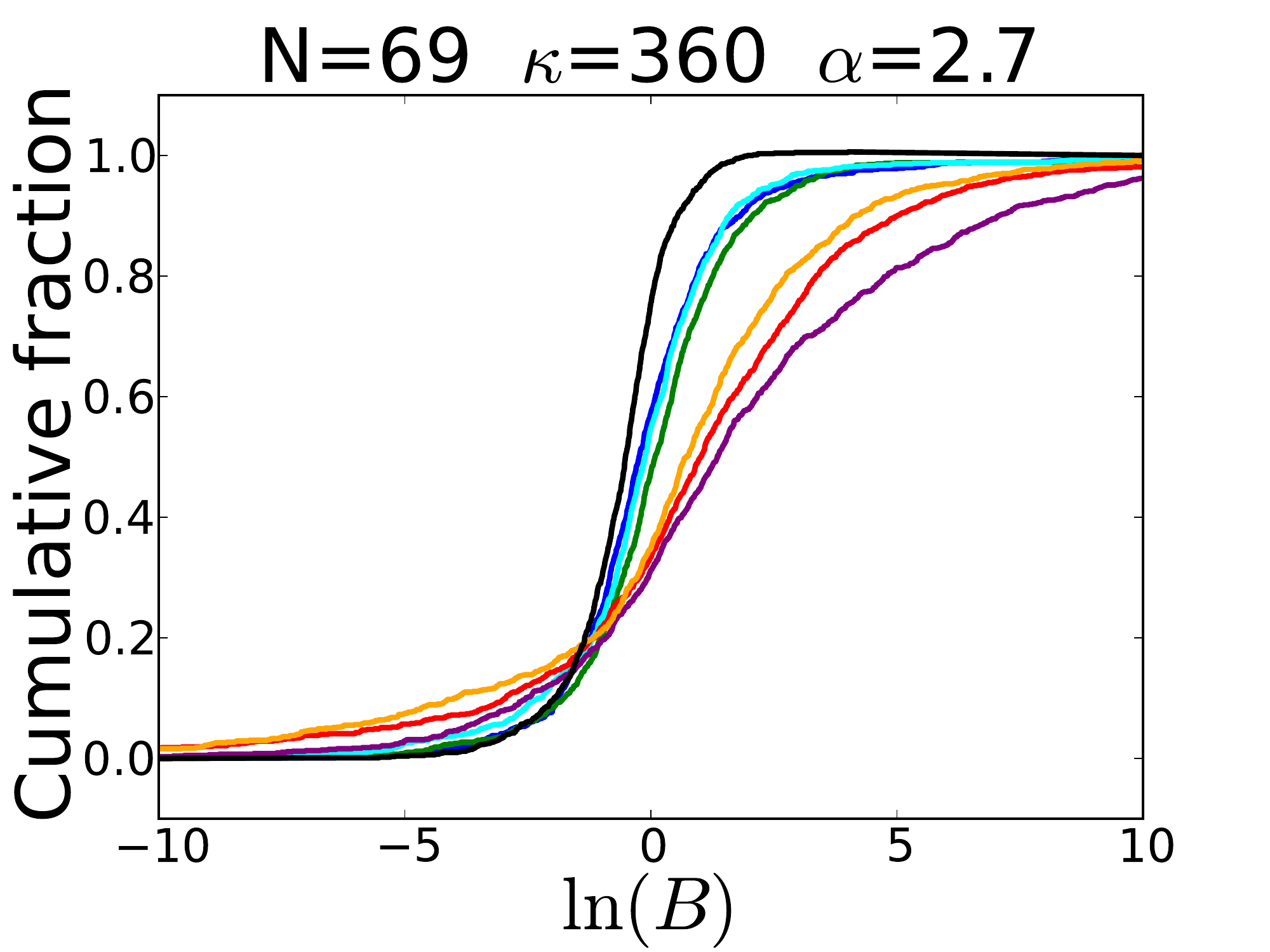} \\
\hrulefill & \hrulefill & \hrulefill
\\
\hbox{\includegraphics[width=40mm]{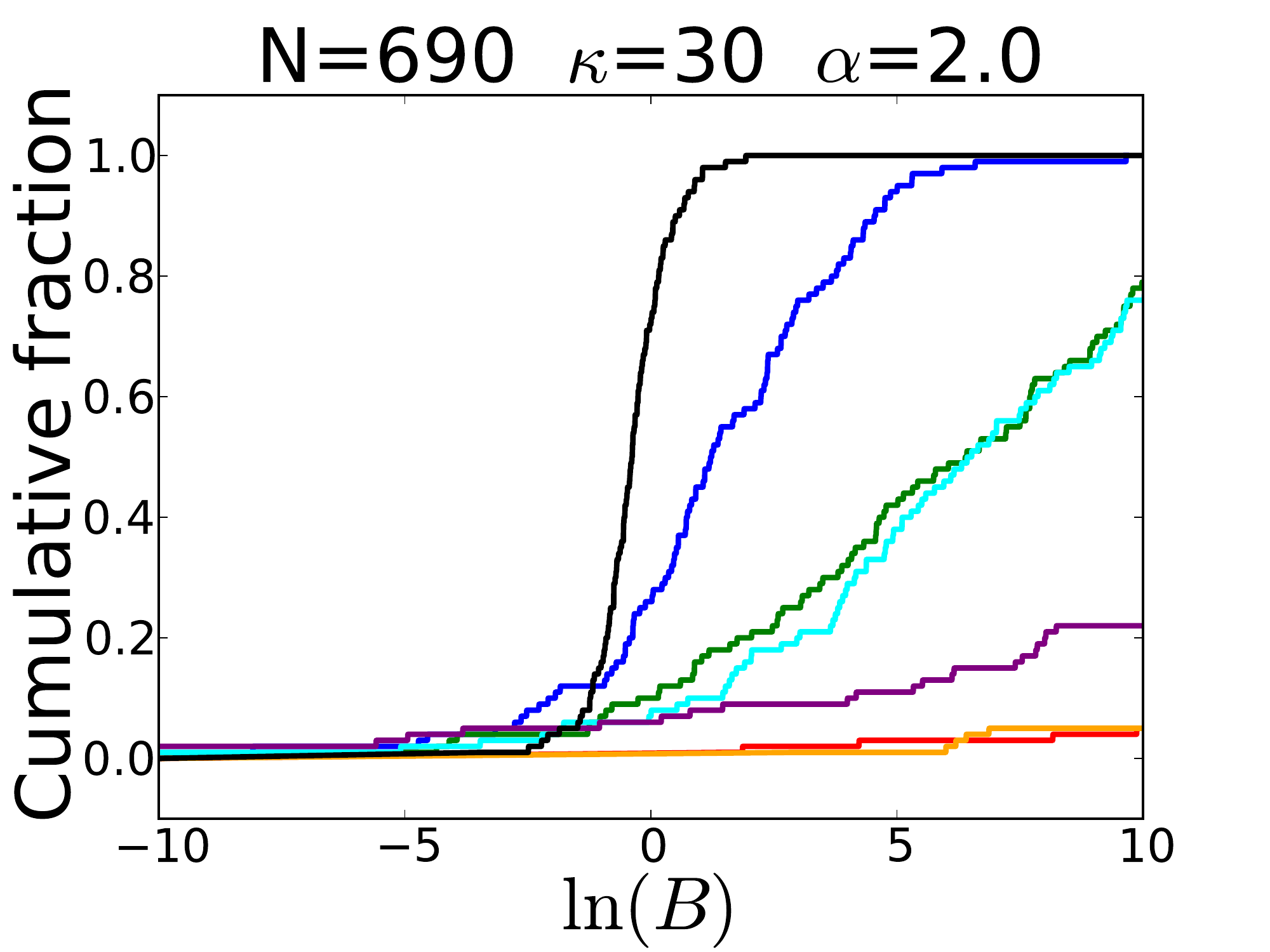}} & \includegraphics[width=40mm]{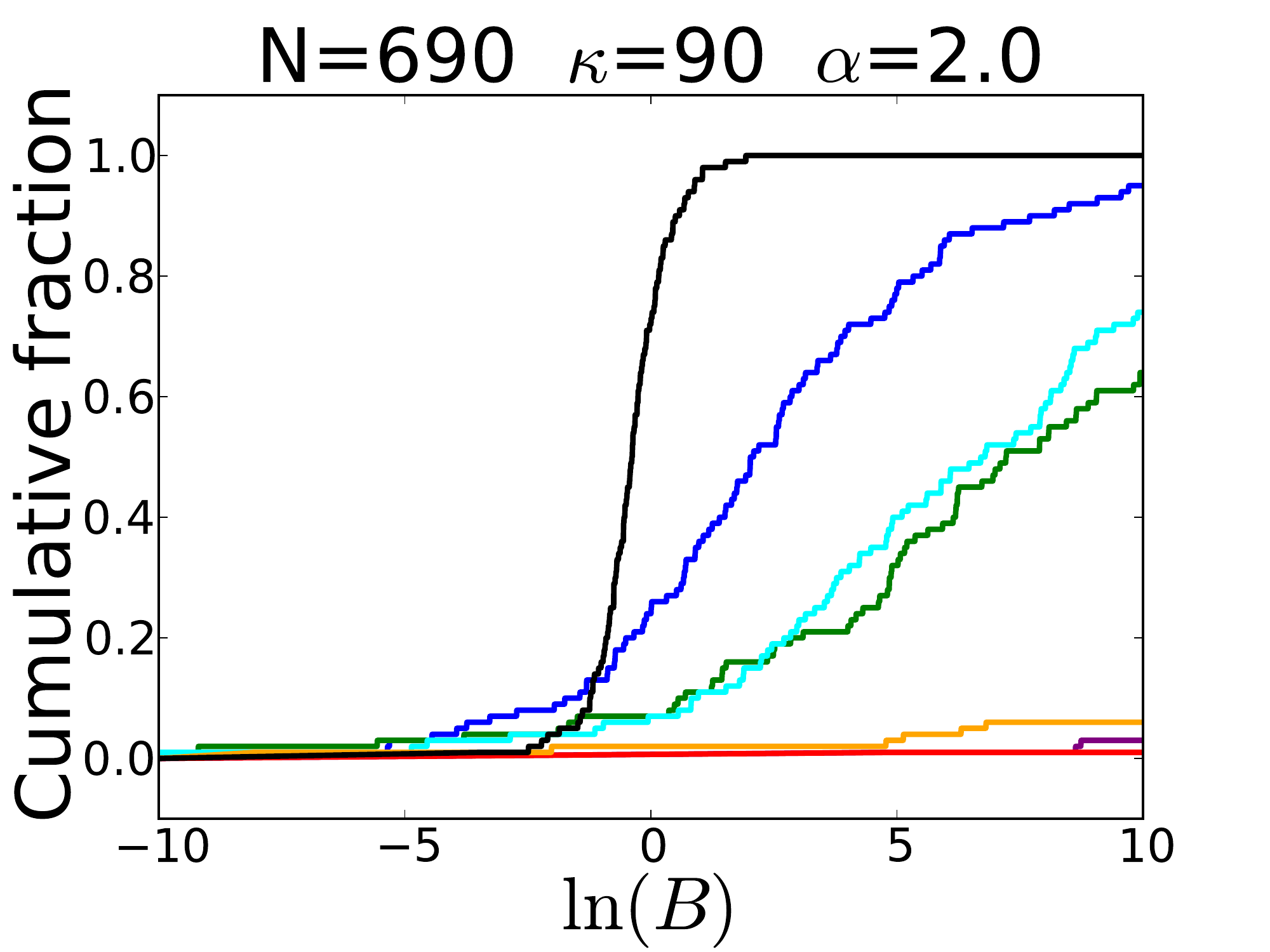} & \includegraphics[width=40mm]{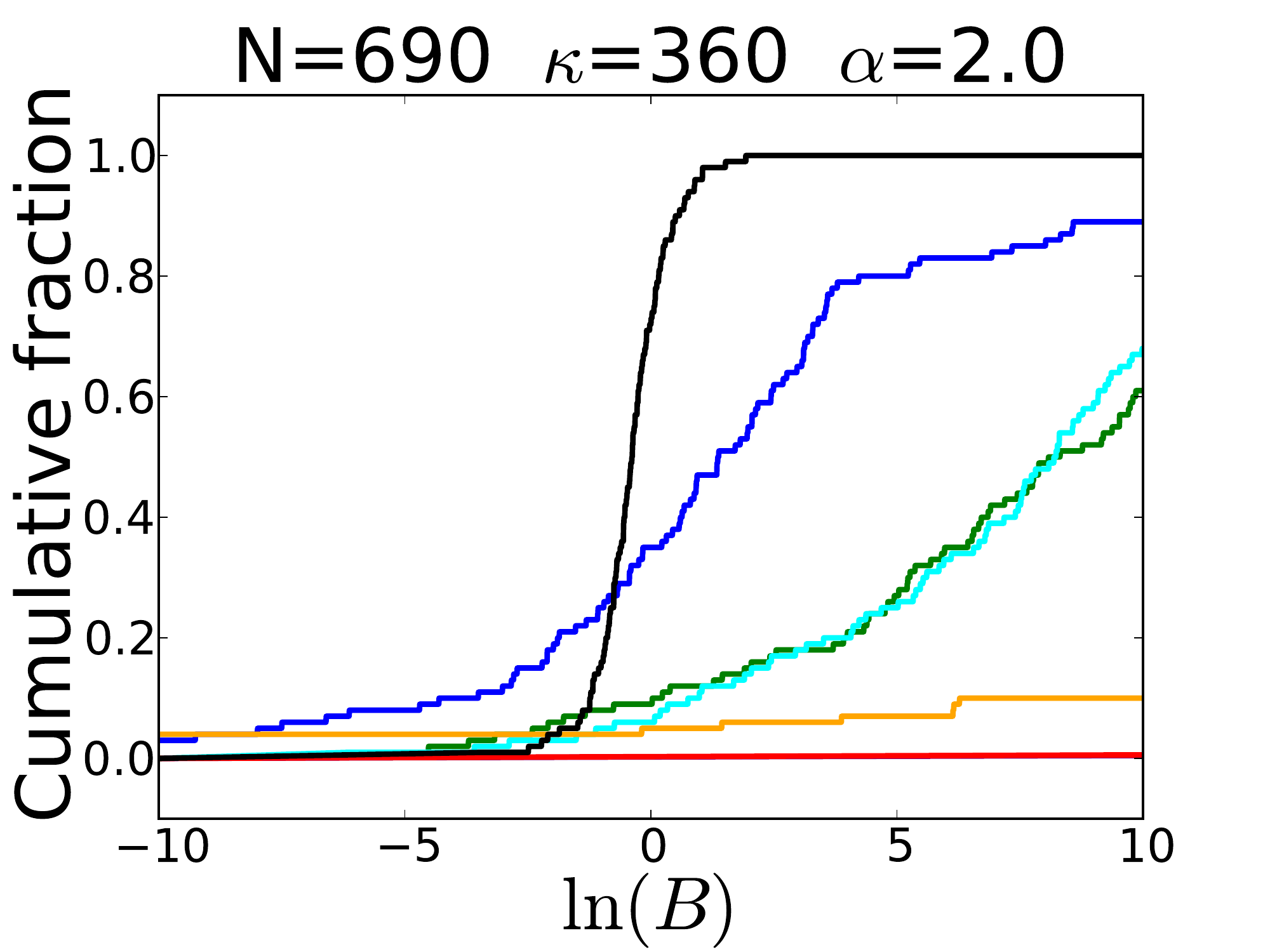} \\
\hbox{\includegraphics[width=40mm]{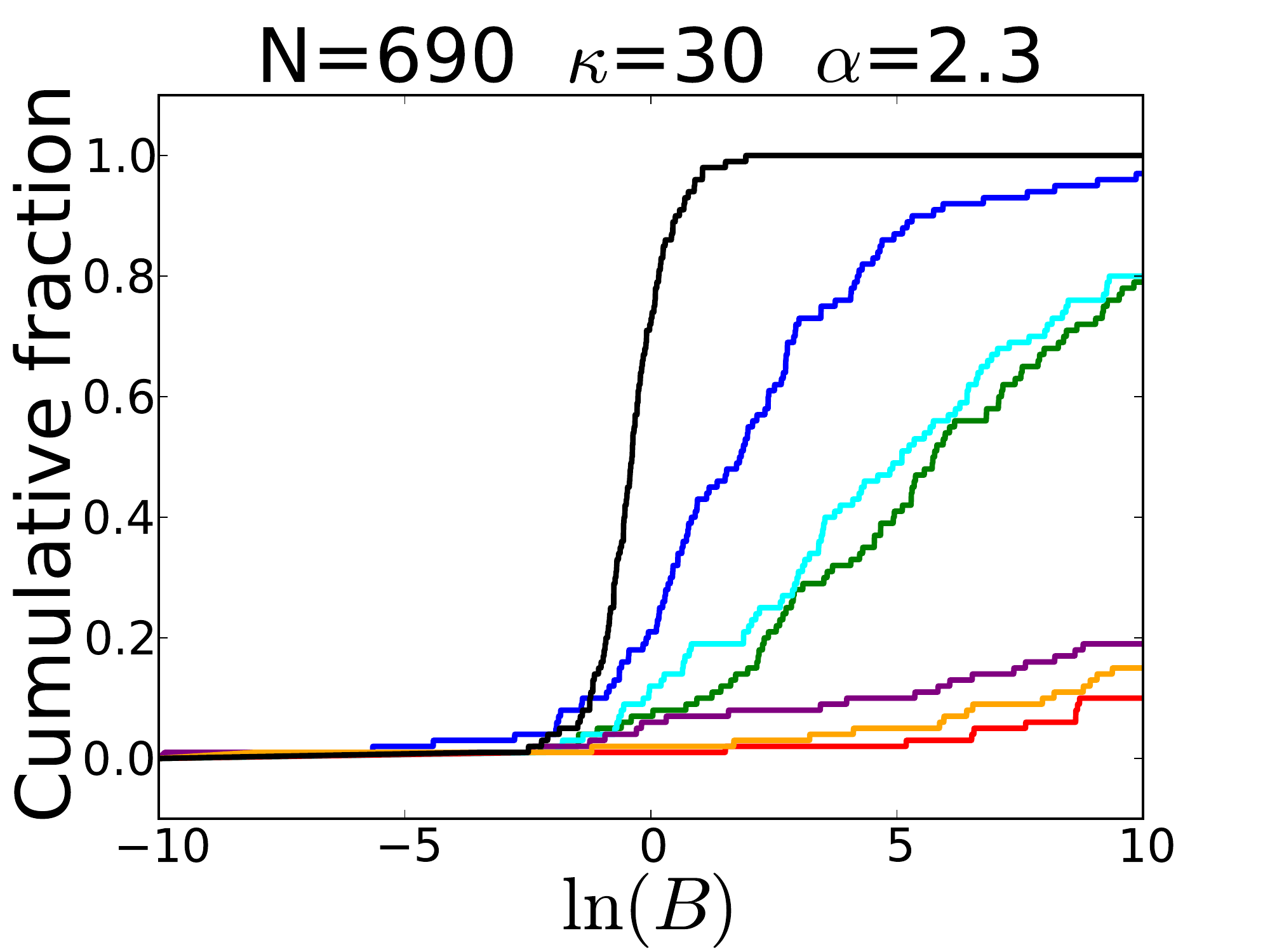}} & \includegraphics[width=40mm]{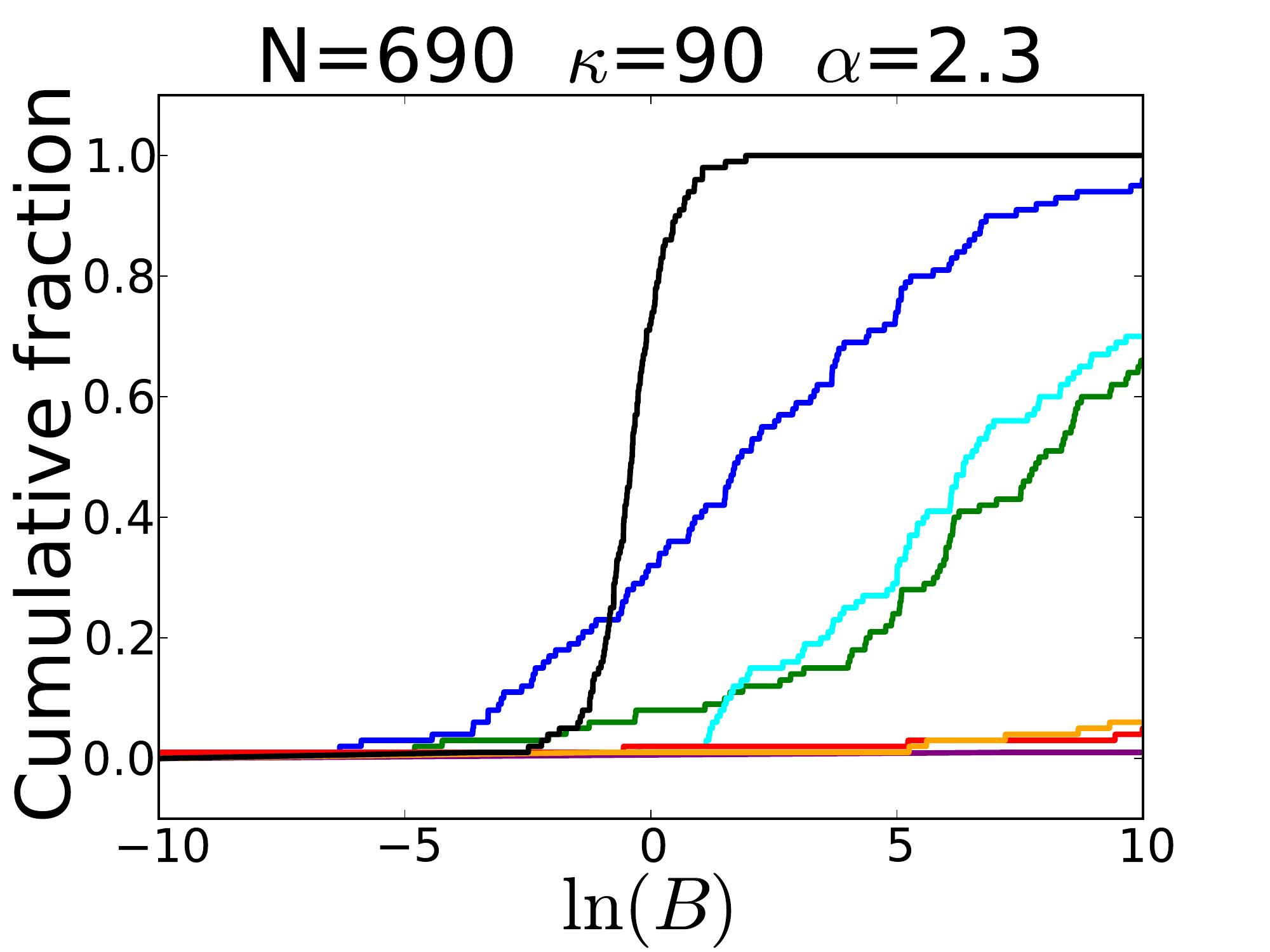} & \includegraphics[width=40mm]{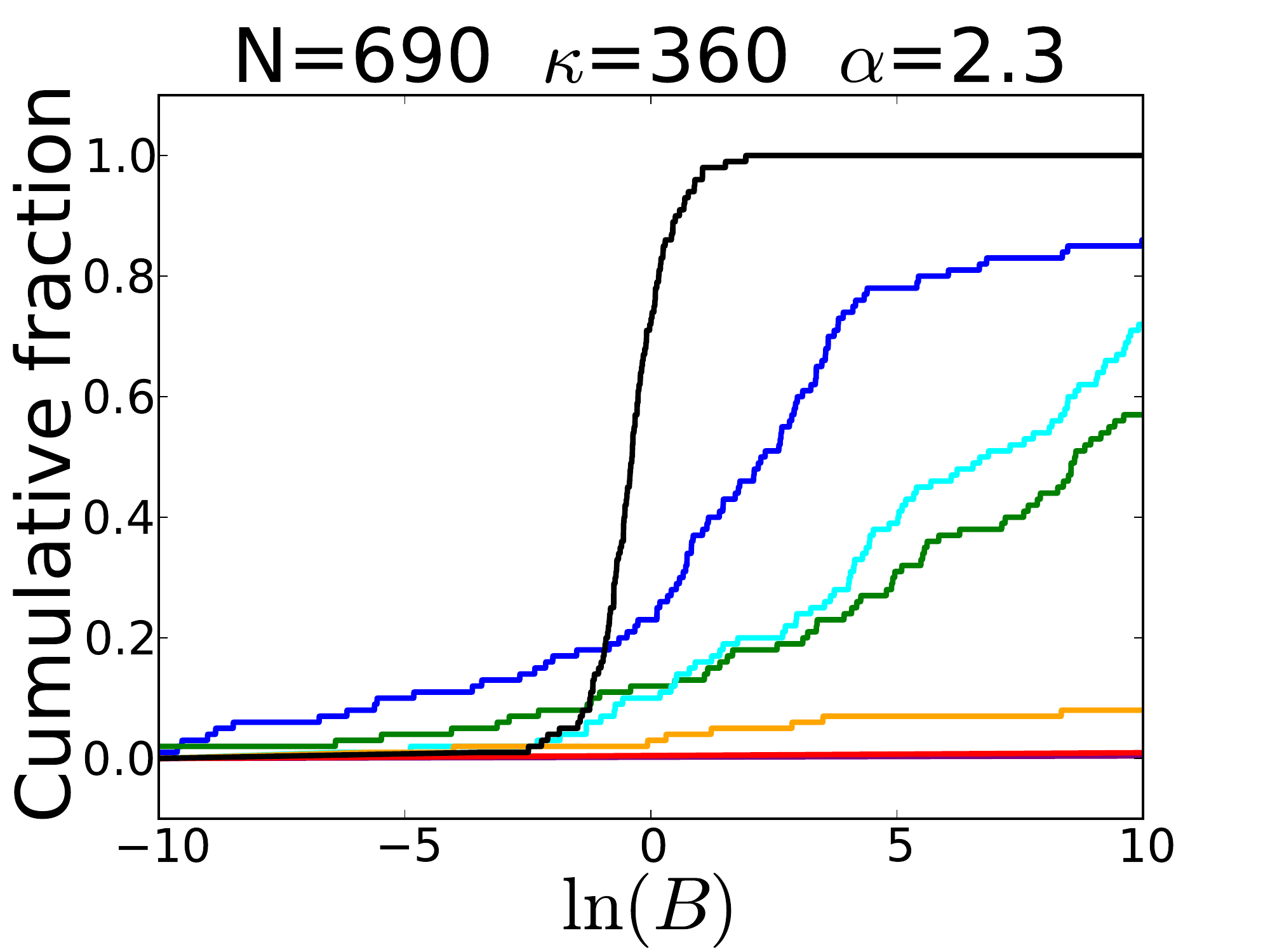} \\
\hbox{\includegraphics[width=40mm]{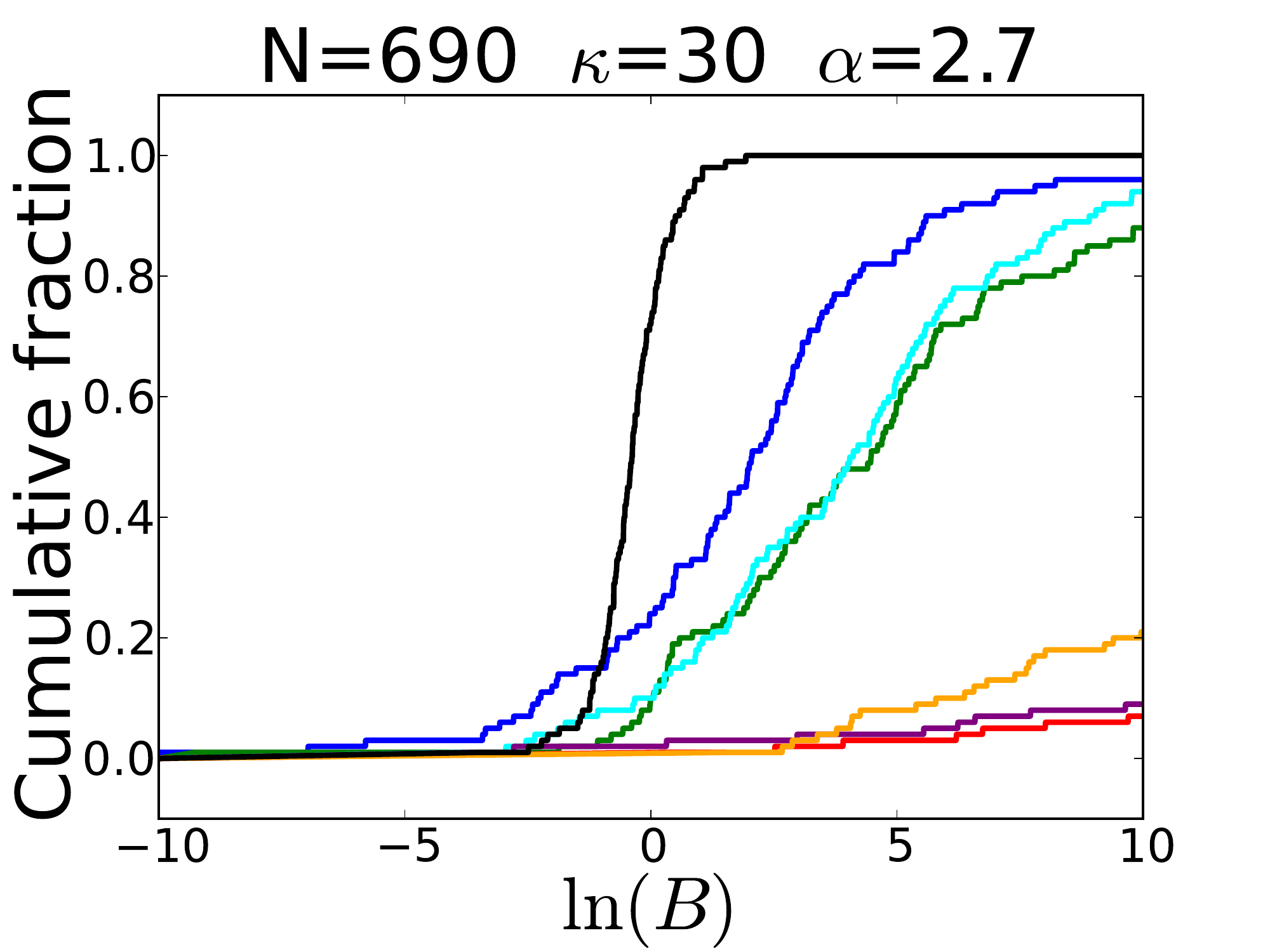}} & \includegraphics[width=40mm]{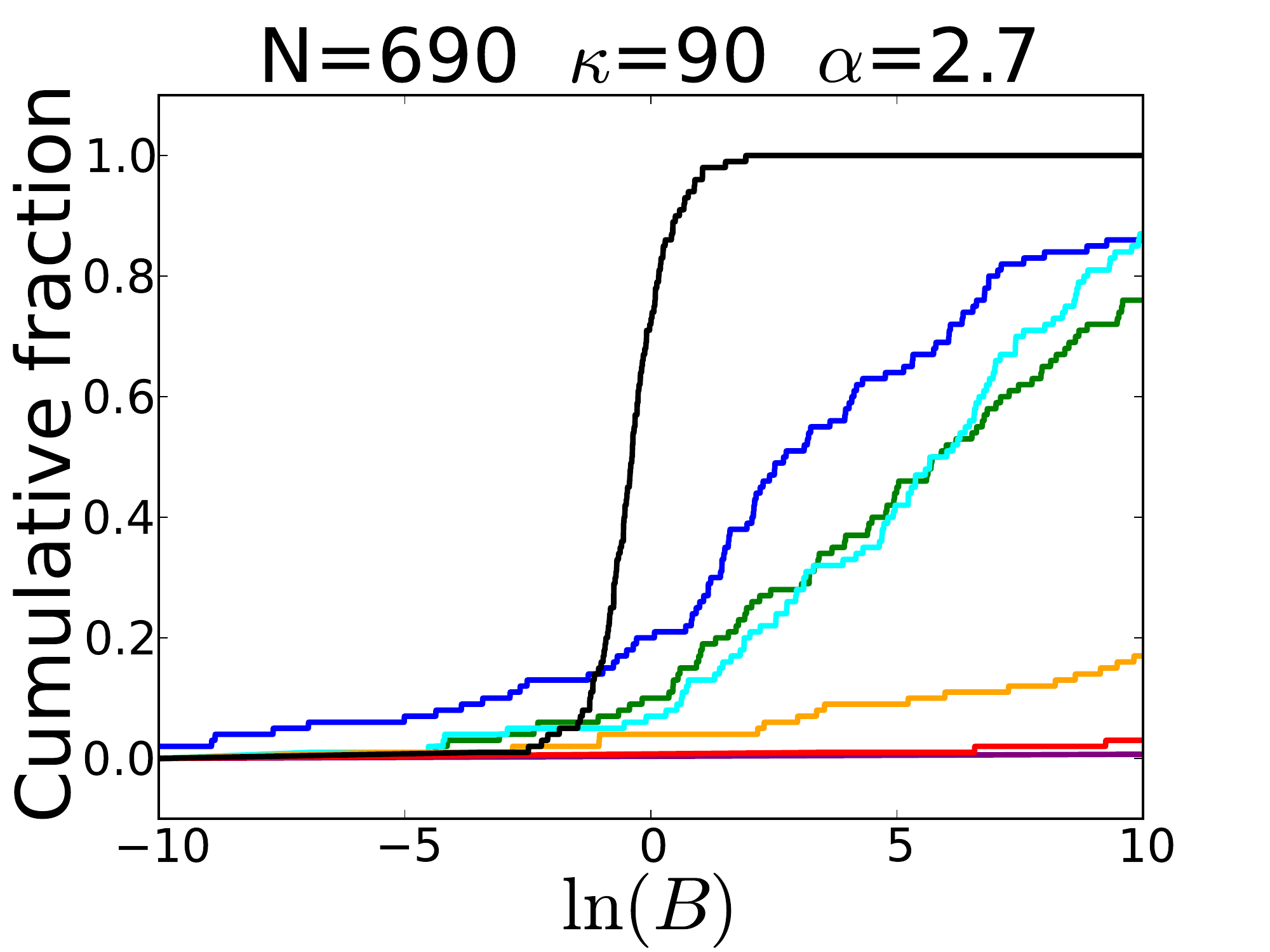} & \includegraphics[width=40mm]{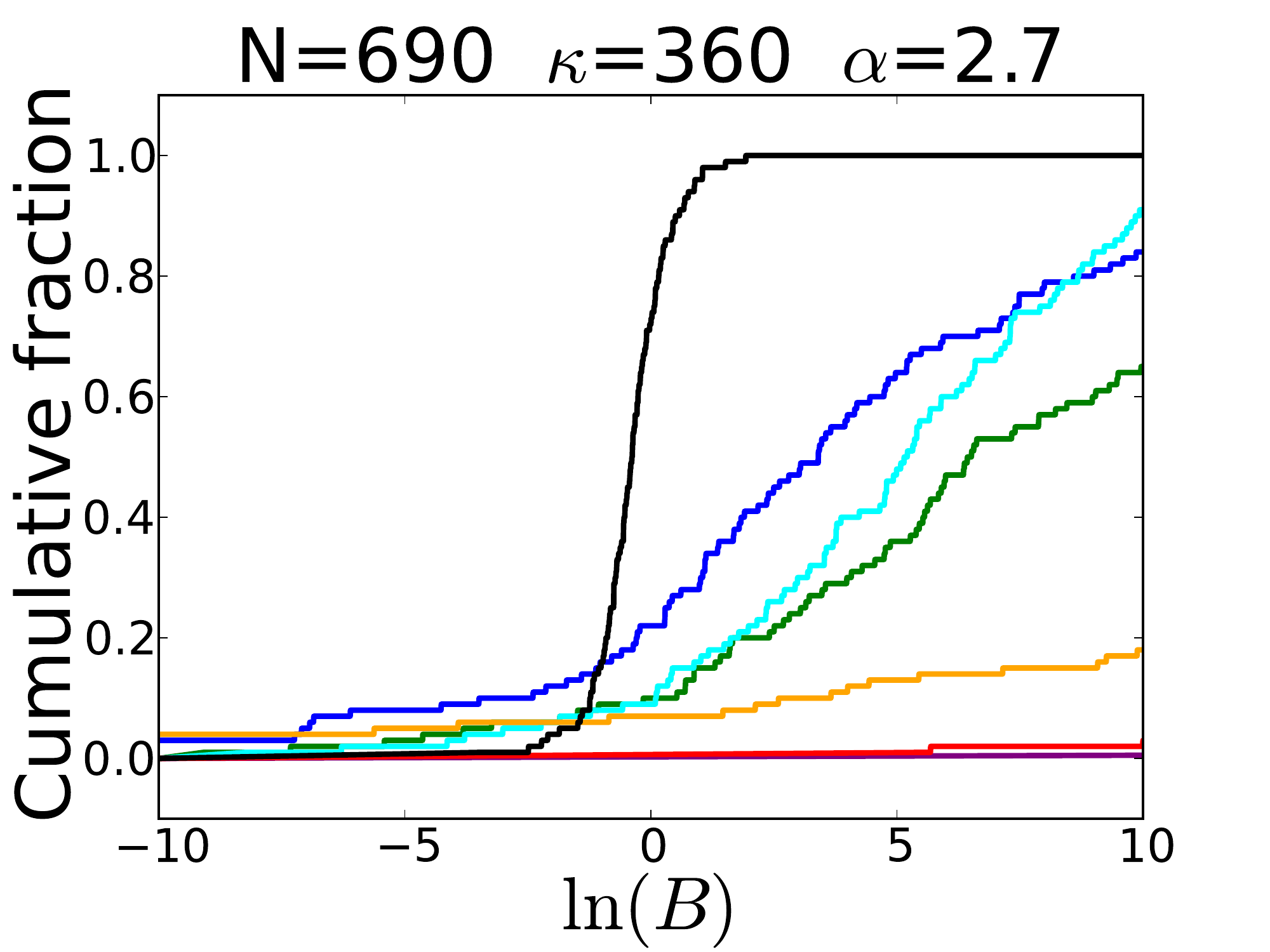}
\end{array}$
\end{center}
\label{fig:A}
\end{figure}
\twocolumn
\clearpage

\section{Analysis of the PAO data}
\label{sec:ApplicationPAO}  

\begin{figure}
\begin{center}$
\arraycolsep=0.01pt\def\arraystretch{0.01}
\begin{array}{c}
\includegraphics[width=90mm]{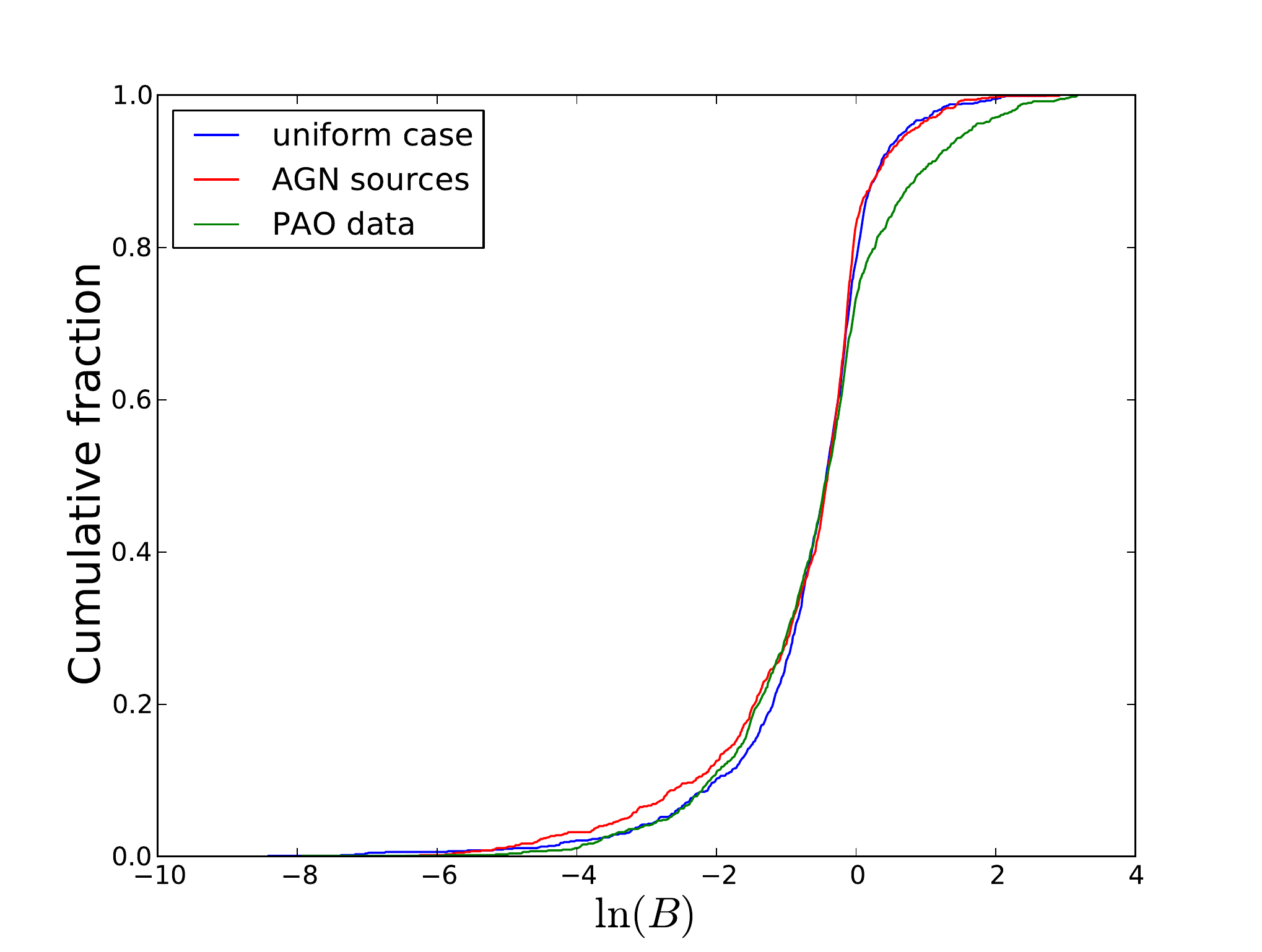}
\end{array}$
\caption{ Cumulative fractions of Bayes factors, produced by the application of the multi-step method to 1,000 partitions of: (a) the PAO data; (b) 69 simulated UHECRs from uniform sources; and (c) 69 simulated UHECRs from a realistic mock catalogue of AGNs. }
\label{fig:1000Bayes}
\end{center}
\end{figure}

We now apply the multi-stage Bayesian method described in Section~\ref{sec:StatMeth} to the PAO data set in order to assess the uniformity of the measured UHECR arrival directions.  This data set consists of 69 events observed from 1 January 2004 to 31 December 2009, and is described in full by \cite{PAO2010}. As the results depend to some extent on the way the data are split into the three subsets, Bayes factors were calculated for 1,000 different random, but equal sized, partitions. The cumulative distribution of Bayes factors is shown in Figure~\ref{fig:1000Bayes}. 

The Bayes factors are calculated for different partitions of the same sample. Apart from the distribution for the PAO data, Figure~\ref{fig:1000Bayes} also shows the distribution for a uniform sample of 69 UHECRs, as well as the distribution for a UHECR sample generated from a realistic AGN catalogue (with a source density of $10^{-3.5}\,{\rm Mpc}^{-3}$, $\kappa=30$ and $\alpha=2.0$). The results shown here differ from those shown in Figure~\ref{fig:AllOfThem}, insofar as they result from different random partitions of a single sample (i.e., PAO, uniform or AGN-sourced) rather than being drawn from completely independent samples.  However, the distributions produced using these two methods are comparable and the main conclusions remain unchanged.

A sensible way of dealing with the range of Bayes factors is to characterize their distribution by the arithmetic or geometric mean.  There is no compelling reason to choose one over the other (see e.g.\ \citealt{OHagan1997}), but the fact that the logarithm of the Bayes factor is symmetric between the two models suggests that the geometric mean is more natural. The geometric mean was 0.57 and the arithmetic mean was 1.26. From Equation~\ref{eq:BayesTheorem}, if we assume a prior probability of 0.5 for both models, we calculate mean posterior probabilities for the clustered model of 0.37 and 0.56 for the respective means. Thus, there is no clear preference for either of the models, and the data are consistent with both. We do not detect evidence for self-clustering. Figure~\ref{fig:1000Bayes} shows that for data sets of this size, the distributions of Bayes factors for the uniform and AGN-centred cases cannot be clearly distinguished. This is consistent with the results of \cite{PAO2012}.
\par
\par

\section{Conclusions} \label{sec:Conc}

We have developed a Bayesian method for the analysis of the self-clustering of points on a sphere and applied it to the 69 highest energy UHECRs detected by PAO up until 31 December 2009.

The method is a three-step Bayesian approach, in which the data are divided into three subsets: the first two subsets of the data are used to generate a model of self-clustered UHECRs; the third subset is used to perform Bayesian model comparison between this self-clustered model and a uniform model of UHECRs. This approach is an extension of the Bayesian model comparison methods that were developed by \cite{Spiegelhalter1982}, \cite{Aitkin1991},  \cite{OHagan1991} and \cite{OHagan1995}. Like the multi-step method that is presented here, those approaches are aimed to evaluate the marginal likelihood in cases when there is weak prior information on the model parameters. The method we have presented here is not specific to the UHECR problem in question and could be applied to anisotropy searches in other areas of astronomy, such as the search for angular anisotropies in the distribution of gamma-ray bursts described by e.g. \cite{Balazs_etal1998} and \cite{Magliocchetti_etal2003}.

There is some ambiguity in the partitioning of the full data set. In the present implementation, the total data set is divided into three subsets of equal size. However, it is possible that a different partitioning, or perhaps an average over partitions could make this method more effective. These issues will be explored in future work.

We tested our model comparison method on mock catalogues of UHECRs. The results for uniform UHECR arrival directions were compared to the results for UHECRs originating in AGNs from a realistic mock catalogue. UHECR clustering in a realistic AGN centred model is too weak to be detected in a sample of 69 events, but would be detectable in samples of 690 events. This is consistent with the results of \cite{PAO2012}.

We assumed a pure proton composition of the cosmic rays, but there are some indications that heavier nuclei are also part of the composition (\citealt{Unger2007}). The effect of including heavier nuclei will be investigated through additional simulations.

For the PAO data, Bayes factors were calculated for different random partitions of the data. The geometric and arithmetic means of the Bayes factors were 0.57 and 1.26 respectively, corresponding to posterior probabilities of 0.37 and 0.56 for the clustered model. Thus, we did not find strong evidence for clustering in the PAO data, although the data are also consistent with the AGN-centred simulations.

It is expected that future experiments will produce data sets that will be sufficiently large for our Bayesian method (and other statistical approaches; see e.g.\ \citealt{FutureAnis2014}) to detect even the weak clustering expected if the UHECRS have come from nearby AGNs. PAO is continuing to take data and is expected to produce a sample of $\sim 250$ UHECRs over its first decade of operations.  Looking further ahead, the planned Japanese Experiment Module Extreme Universe Space Observatory (JEM-EUSO, \citealt{JEMEUSO2013}) on the International Space Station (ISS) is scheduled for launch in 2017 and is expected to detect $\sim 200$ UHECRs annually over its five year lifetime. These data sets should be sufficiently large to detect the self-clustering of UHECRs independent of the source population.



\section*{Acknowledgments}

We thank Andreas Berlind and Glennys Farrar for making their mock catalogues public and Todor Stanev for providing the results of his UHECR propagation models. We thank an anonymous referee for making suggestions which allowed us to improve this paper. AK was supported by a Science \& Technology Facilities Council studentship. 
 

\bibliographystyle{mn2e}
\bibliography{xxx}


\bsp
\label{lastpage}
\end{document}